\PassOptionsToPackage{unicode}{hyperref}
\PassOptionsToPackage{hyphens}{url}
\PassOptionsToPackage{dvipsnames,svgnames,x11names}{xcolor}
\documentclass[
  12pt,
  number]{elsarticle}
\usepackage{xcolor}
\usepackage[inner=1in,outer=1in,top=1in,bottom=1in]{geometry}
\usepackage{amsmath,amssymb}
\usepackage{iftex}
\ifPDFTeX
  \usepackage[T1]{fontenc}
  \usepackage[utf8]{inputenc}
  \usepackage{textcomp} 
\else 
  \usepackage{unicode-math} 
  \defaultfontfeatures{Scale=MatchLowercase}
  \defaultfontfeatures[\rmfamily]{Ligatures=TeX,Scale=1}
\fi
\usepackage{lmodern}
\ifPDFTeX\else
\fi
\IfFileExists{upquote.sty}{\usepackage{upquote}}{}
\IfFileExists{microtype.sty}{
  \usepackage[]{microtype}
  \UseMicrotypeSet[protrusion]{basicmath} 
}{}
\usepackage{setspace}
\makeatletter
\@ifundefined{KOMAClassName}{
  \IfFileExists{parskip.sty}{%
    \usepackage{parskip}
  }{
    \setlength{\parindent}{0pt}
    \setlength{\parskip}{6pt plus 2pt minus 1pt}}
}{
  \KOMAoptions{parskip=half}}
\makeatother
\makeatletter
\ifx\paragraph\undefined\else
  \let\oldparagraph\paragraph
  \renewcommand{\paragraph}{
    \@ifstar
      \xxxParagraphStar
      \xxxParagraphNoStar
  }
  \newcommand{\xxxParagraphStar}[1]{\oldparagraph*{#1}\mbox{}}
  \newcommand{\xxxParagraphNoStar}[1]{\oldparagraph{#1}\mbox{}}
\fi
\ifx\subparagraph\undefined\else
  \let\oldsubparagraph\subparagraph
  \renewcommand{\subparagraph}{
    \@ifstar
      \xxxSubParagraphStar
      \xxxSubParagraphNoStar
  }
  \newcommand{\xxxSubParagraphStar}[1]{\oldsubparagraph*{#1}\mbox{}}
  \newcommand{\xxxSubParagraphNoStar}[1]{\oldsubparagraph{#1}\mbox{}}
\fi
\makeatother

\usepackage{longtable,booktabs,array}
\usepackage{calc} 
\usepackage{etoolbox}
\makeatletter
\patchcmd\longtable{\par}{\if@noskipsec\mbox{}\fi\par}{}{}
\makeatother
\IfFileExists{footnotehyper.sty}{\usepackage{footnotehyper}}{\usepackage{footnote}}
\makesavenoteenv{longtable}
\usepackage{graphicx}
\makeatletter
\newsavebox\pandoc@box
\newcommand*\pandocbounded[1]{
  \sbox\pandoc@box{#1}%
  \Gscale@div\@tempa{\textheight}{\dimexpr\ht\pandoc@box+\dp\pandoc@box\relax}%
  \Gscale@div\@tempb{\linewidth}{\wd\pandoc@box}%
  \ifdim\@tempb\p@<\@tempa\p@\let\@tempa\@tempb\fi
  \ifdim\@tempa\p@<\p@\scalebox{\@tempa}{\usebox\pandoc@box}%
  \else\usebox{\pandoc@box}%
  \fi%
}
\def\fps@figure{htbp}
\makeatother

\ifLuaTeX
  \usepackage{luacolor}
  \usepackage[soul]{lua-ul}
\else
  \usepackage{soul}
\fi

\NewDocumentCommand\citeproctext{}{}

\makeatletter
 \let\@cite@ofmt\@firstofone
 \def\@biblabel#1{}
 \def\@cite#1#2{{#1\if@tempswa , #2\fi}}
\makeatother
\newlength{\cslhangindent}
\newlength{\csllabelwidth}
\newenvironment{CSLReferences}[2] 
 {\begin{list}{}{%
  \setlength{\itemindent}{0pt}
  \setlength{\leftmargin}{0pt}
  \setlength{\parsep}{0pt}
  \ifodd #1
   \setlength{\leftmargin}{\cslhangindent}
   \setlength{\itemindent}{-1\cslhangindent}
  \fi
  \setlength{\itemsep}{#2\baselineskip}}}
 {\end{list}}
\usepackage{calc}

\newcommand{\CSLLeftMargin}[1]{\parbox[t]{\csllabelwidth}{\strut#1\strut}}
\newcommand{\CSLRightInline}[1]{\parbox[t]{\linewidth - \csllabelwidth}{\strut#1\strut}}

\makeatletter
\@ifpackageloaded{caption}{}{\usepackage{caption}}
\AtBeginDocument{%
\ifdefined\contentsname
  \renewcommand*\contentsname{Table of contents}
\else
  \newcommand\contentsname{Table of contents}
\fi
\ifdefined\listfigurename
  \renewcommand*\listfigurename{List of Figures}
\else
  \newcommand\listfigurename{List of Figures}
\fi
\ifdefined\listtablename
  \renewcommand*\listtablename{List of Tables}
\else
  \newcommand\listtablename{List of Tables}
\fi
\ifdefined\figurename
  \renewcommand*\figurename{Figure}
\else
  \newcommand\figurename{Figure}
\fi
\ifdefined\tablename
  \renewcommand*\tablename{Table}
\else
  \newcommand\tablename{Table}
\fi
}
\@ifpackageloaded{float}{}{\usepackage{float}}
\floatstyle{ruled}
\@ifundefined{c@chapter}{\newfloat{codelisting}{h}{lop}}{\newfloat{codelisting}{h}{lop}[chapter]}
\floatname{codelisting}{Listing}

\makeatother
\makeatletter
\@ifpackageloaded{caption}{}{\usepackage{caption}}
\@ifpackageloaded{subcaption}{}{\usepackage{subcaption}}
\makeatother
\usepackage{bookmark}
\IfFileExists{xurl.sty}{\usepackage{xurl}}{} 
\hypersetup{
  pdftitle={Time-to-Event Estimation with Unreliably Reported Events in Medicare Health Plan Payment},
  pdfauthor={Oana M. Enache; Sherri Rose},
  pdfkeywords={survival analysis, restricted mean time lost, restricted
mean survival time, Medicare Advantage, upcoding, risk adjustment},
  colorlinks=true,
  linkcolor={blue},
  filecolor={Maroon},
  citecolor={Blue},
  urlcolor={Blue},
  pdfcreator={LaTeX via pandoc}}

\begin{document}

\begin{frontmatter}
\title{Time-to-Event Estimation with Unreliably Reported Events in
Medicare Health Plan Payment}
\author[1]{Oana M. Enache%
\corref{cor1}%
}
 \ead{oenache@stanford.edu} 
\author[2]{Sherri Rose%
}

\affiliation[1]{organization={Stanford University School of
Medicine, Department of Biomedical Data Science},addressline={Edwards
Building, 300 Pasteur Drive},city={Stanford,
CA},postcode={94304},postcodesep={}}
\affiliation[2]{organization={Stanford University, Department of Health
Policy},addressline={Encina Commons, 615 Crothers Way},city={Stanford,
CA},postcode={94305},postcodesep={}}

\cortext[cor1]{Corresponding author}

\begin{abstract}
\textbf{Objective}: To propose time-to-event estimators that help
evaluate incident diagnostic coding and possible upcoding in Medicare as
well as introduce an open-source software package that enables more
reproducible methods development relevant to Medicare billing
behavior.\\
\textbf{Study setting and Design}: Observational analysis of simulated
upcoding based on coding by insurers or providers that may be
incentivized by Medicare Advantage risk adjustment.\\
\textbf{Data Sources and Analytic Sample}: Two years of separately
simulated incident health condition coding data for a Medicare Advantage
population and a Traditional Medicare population where coding patterns
are aligned with known practices in each program.\\
\textbf{Principal Findings}: We propose several novel time-to-event
estimators of incident coding intensity and possible upcoding in
Medicare Advantage, including accounting for unreliable reporting. We
demonstrate estimator performance in simulated data leveraging the
National Institutes of Health's All of Us study and also develop an open
source R package to simulate longitudinal realistic labeled upcoding
data, which were not previously available for researchers. In
simulations, our novel estimators recovered differences in upcoding
within and across monitoring periods. Undercoding had a limited effect
on our novel estimators while an existing estimator was more sensitive
to undercoding.\\
\textbf{Conclusions}: Our proposed estimators can help researchers and
policymakers track new coding behaviors (e.g., as may be incentivized by
risk adjustment formula updates) earlier and at scale while accounting
for several real-world data considerations. Further, the R package we
provide can be used to improve the development, accessibility, and
reproducible evaluation of coding intensity and upcoding methodology.
\end{abstract}

\end{frontmatter}

\setstretch{1}
\newpage{}

\ul{\textbf{Callout Box}}

\textbf{What is known on this topic:}

\begin{itemize}
\item
  The present structure of health plan payment in Medicare Advantage
  incentivizes high coding intensity and possible upcoding, costing the
  federal government billions annually without clear benefit to
  beneficiaries.
\item
  No individual-level labeled upcoding data are currently available to
  researchers, making developing and evaluating coding intensity and
  upcoding measures and estimators challenging.
\item
  Current coding intensity estimation approaches largely count diagnoses
  within a fixed time period (e.g., a year) and do not account for
  issues like Traditional Medicare undercoding.
\end{itemize}

\textbf{What this study found:}

\begin{itemize}
\item
  This study introduces a set of time-to-event estimators of coding
  intensity that address some unrealistic assumptions of widely-used
  existing estimators.
\item
  Providing an open-source R package enables the simulation of realistic
  baseline, undercoded, and upcoded data over time, including labels for
  upcoding instances.
\item
  The proposed estimators and software package can help researchers
  better evaluate Medicare providers and insurers' coding behaviors and
  identify health plan payment improvements at scale.
\end{itemize}

\section{Introduction}\label{introduction}

Medicare is a federally funded insurance program administered by the
Centers for Medicare and Medicaid Services (CMS), supporting over 69
million Americans who are aged 65 years and older or chronically
disabled\textsuperscript{1}. Beneficiaries choose between receiving
coverage through Traditional Medicare (TM) or Medicare Advantage (MA).
TM beneficiaries' care is generally paid directly by CMS for each
service provided. In contrast, MA beneficiaries receive health insurance
from private insurers who are paid by CMS for each beneficiary's
anticipated health needs.

These prospective health plan payments are determined by risk
adjustment, currently an ordinary least squares linear regression with
binary demographic and health diagnostic variables, each of which has a
positive coefficient. The outcome of this algorithm is a risk score,
which is used to adjust a benchmark payment amount per
beneficiary\textsuperscript{2,3}. There are 115 variables (hierarchical
condition categories or HCCs) that report the purported presence or
absence of certain health conditions in the formula\textsuperscript{4},
which is also referred to as CMS-HCC\textsuperscript{3}. Although HCCs
overall correspond to dozens of distinct health conditions, certain
subsets of these variables also represent severity levels within the
same condition and are therefore billed in a mutually exclusive manner,
where payment increases with severity. For example, in the current
version (version 28 or V28) of CMS-HCC an insurer can only be paid for
coding one of HCC125-127, which correspond to severe, moderate, and mild
or unspecified dementia respectively\textsuperscript{3,4}.

Importantly, insurers retain the amount of money paid by CMS regardless
of what care their beneficiaries actually receive\textsuperscript{2,3}.
This and the structure of CMS-HCC incentivizes insurers to code for as
many diagnoses as possible to increase profits\textsuperscript{5--7}.
Insurers coding for more diagnoses or more severe diagnoses than is
accurate is referred to as upcoding. This is thought to cost CMS tens of
billions of dollars annually in unnecessary spending to private insurers
without clear benefit to MA beneficiaries\textsuperscript{7}. Most major
MA insurers have also been accused of fraud due to billing practices at
least once in recent years by the United States Department of Justice,
largely using information from whistleblowers or other manual
audits\textsuperscript{8--11}.

Prior literature has also estimated upcoding inconsistently for several
decades\textsuperscript{6}, in part because there are a number of
barriers for researchers working on such methods. First, it is difficult
to definitively identify upcoding when researchers only have access to
national Medicare claims data, as is typical. Second, gaining access to
individual-level Medicare data is a time-consuming and sometimes
inaccessible process. Third, there are limited national data resources
describing co-occurring health conditions free of coding incentives in
older adults\textsuperscript{12--14}. Fourth, labeled upcoding data to
evaluate estimators are not available. This also means that comparing
proposed methods is challenging, as the data used to develop or evaluate
methods are often not able to be shared.

Thus, assessment of upcoding in MA is usually done by comparing
prevalence of coding of beneficiaries, or coding intensity, to coding
intensity of similar health conditions in TM, with slight variation on
inclusion and exclusion criteria used\textsuperscript{15}. However, TM
is known to have underreporting (i.e., undercoding) of health
conditions\textsuperscript{3,16}. This type of unreliable reporting is
important to account for as measures of differences between the programs
will likely be inflated otherwise. Alternative reference data besides TM
diagnoses have been explored in prior work\textsuperscript{17,18} but
remain underutilized.

To our knowledge, time-to-event (TTE) analyses (i.e., survival analysis)
have not previously been used to evaluate coding behaviors in Medicare.
TTE analyses typically aim to compare differences between groups drawn
from two or more populations in a randomized clinical trial or
observational study with censoring, one type of incompletely observed
data. This may include plotting events over time using either
Kaplan-Meier or cumulative incidence curves\textsuperscript{19,20}.

Most commonly, hazard ratios are used to estimate the magnitude of the
effect between groups\textsuperscript{21--23}. For example, a review of
66 clinical trials with TTE primary outcomes across four major medical
journals found that 80\% of trials reported a hazard ratio as a main
finding, while only 21\% of studies reported any alternative
approaches\textsuperscript{24}. However, hazard ratios are unintuitive
to interpret\textsuperscript{22,25} and rely on an often unrealistic
proportional hazards assumption\textsuperscript{21,24,26}. Researchers
sometimes also compare mean or median survival, but this omits most of
the data and may not be estimable in certain
scenarios\textsuperscript{22,27}.

Furthermore, alternate approaches are needed if competing risks, or when
there is more than one mutually exclusive outcome, are present. Despite
being pervasive in health studies, competing risks are frequently
ignored, which can result in biased estimates of the primary
outcome\textsuperscript{20,28,29}. Additional important considerations
in TTE analyses include the choice of monitoring period (i.e., the time
window analyzed between a pre-specified origin and end time) and
comparison group.

Although first proposed in 1949\textsuperscript{30}, restricted mean
survival time (RMST) was revisited for TTE estimation in more
contemporary literature\textsuperscript{25,26,31--33} because it
addresses many of the issues of more popular approaches. RMST is defined
as the area under the survival curve of time to an event for a single
monitoring period. It can be interpreted as the mean time to event for
all study participants followed in that monitoring
period\textsuperscript{25}. With large enough sample sizes, it is
estimable nonparametrically, does not require proportional hazards, and
is censoring independent\textsuperscript{25}.

In TTE analyses with a single outcome, the area above the survival curve
in one monitoring period (or, the end time minus the RMST) is the
restricted mean time lost (RMTL)\textsuperscript{34}. RMTL has a number
of appealing features, including that it can straightforwardly be
extended to correspond to the area below cause- or event-specific
cumulative incidence curves in competing risk
settings\textsuperscript{29,34,35}. In such scenarios, an event-specific
RMTL corresponds to the area below its cumulative incidence curve and
can be interpreted as the mean time without that particular event in a
monitoring period\textsuperscript{29}, or a summary of both how many and
when events occur in that time. Differences in RMTL between two groups
(for example, a treatment and control or other comparison group) can
also be used to quantify effects\textsuperscript{36}.

We expand RMTL estimation approaches for use in evaluating incident HCC
coding in MA following CMS-HCC formula updates. Here, individual HCCs
and sets of mutually exclusive HCCs corresponding to severity levels of
a single health condition are our events of interest. We propose that
the latter HCC type can be considered conceptually analogous to
competing risks in TTE analyses because only one HCC can be recorded and
paid for at a time. An event occurs when an a diagnosis not reported in
prior monitoring periods is reported by an insurer.

Besides being a unique application of TTE methods in itself, our
approach expands on prior RMTL-based methodological work by also
accounting for potential underreporting of events in TM. Further, we
propose a novel estimation approach for identifying one type of possible
severity-based upcoding, by which we mean beneficiaries who have a
lower-severity version of an HCC being coded with the most severe HCC of
that health condition. Other estimators also apply to a different form
of upcoding that we call any-available, where any beneficiaries
previously not coded with a given HCC could be reported as having that
health condition, potentially fraudulently. When upcoding has not been
concretely confirmed (as is most often the case in real-world settings)
we reflect this by describing it as possible upcoding.

Our contributions also include the creation of an R package to simulate
co-occurring HCCs based on older American adults' self-reported health
conditions from a large national survey. Self-reported data are not
impacted by coding incentives to the same degree as claims or electronic
health record data\textsuperscript{12--14} and enable analysis of
incident coding without underreporting. Our package additionally allows
users to modify these baseline data by either underreporting existing
baseline diagnoses (realistic to TM data) or upcoding specific HCCs over
time, which are then labeled. Labeled upcoding data did not previously
exist---existing individual-level Medicare do not contain this
information---so these simulated data are broadly useful for methods
development in health policy.

\section{Methods}\label{methods}

\subsection{Estimation approach}\label{estimation-approach}

In our setting, TTE and time to reported event are equivalent because we
do not observe each coding event directly; because our estimates are
over entire monitoring periods, we do not consider potential delays in
reporting events to have notable impact. Multiple events may be reported
at each of several time points within one or more monitoring periods
(e.g., a monitoring period could be a calendar year where each time
interval is three months). If there are multiple monitoring periods
(e.g., years) where events are reported, then each monitoring period has
a separate origin and end time between which events are reported. Our
goals are to both (1) evaluate reporting of incident events within one
monitoring period and (2) compare incident event reporting across
sequential monitoring periods.

\subsubsection{Notation for incident
events}\label{notation-for-incident-events}

An event (reported coding of either a single HCC or a single member of a
set of HCCs corresponding to severity levels of one health condition) is
considered incident if it was not reported in prior monitoring periods.
The reporting of an incident event is represented by a vector \(S\),
with \(s \ge 1\) possible mutually exclusive subtype events. \(S\)
encodes which of these events occurs first, analogous to competing
risks. These are referred to as competing events and examined in a
event-specific manner. When \(s > 1\), the possible values \(S\) can
take are written \(s \in \{1, ..., k\}\) in order of increasing
severity. Incident reporting of these events is observed in a set of two
independent groups labeled by \(g\), where \(g=0\) is a comparison group
for \(g=1\), over \(m \ge 2\) pre-specified monitoring periods.

For a given monitoring period, group, and subtype event, \(T\) is the
true time to incident reporting for that event only. In addition, \(C\)
is the time to censoring not due to any competing event. We observe
\(Y= \text{min}(T, C)\) and know whether censoring or reporting of the
event happened first, which is denoted by \(\Delta = I(T \le C)\). Our
observed data are therefore of the form
\(\{(Y_1, S_1\Delta_1),...,(Y_n, S_n\Delta_n)\}\) for the \(n\) total
events reported within the monitoring period. Ordered discrete event
reporting times are \(t_1 < t_2 < ... < \tau\), where \(\tau\) is the
end time of the monitoring period. A single time of event reporting in a
monitoring period is denoted \(t_i\). Finally, when comparing across
groups or monitoring periods we add a respective \(g\) or \(m\)
subscript to the notation above. Sequential monitoring periods are
written as \(m-1\) and \(m\).

\subsubsection{Notation for reference
events}\label{notation-for-reference-events}

We use a set of reference events, or HCCs distinct from the HCCs
examined for incident reporting, to estimate underreporting as others
have done previously\textsuperscript{3,16}. In theory, reference HCCs
reported in one monitoring period should continue to be reported at
equal rates in subsequent monitoring periods. However, in practice some
reference events (and events in \(g=0\) more broadly) may be
underreported. Reference events are denoted as a vector \(S^*\) of \(h\)
distinct HCCs, which can take values \(s^* \in \{1, ..., h\}\). None of
these reference events have competing events. The persistence of a given
reference event \(s^*\) in monitoring period \(m\), meaning the
proportion of individuals coded with \(s^*\) in monitoring period \(m\)
who were previously coded with \(s^*\) in monitoring period \(m-1\), is
\(q_{s^*,m}\) in line with prior work\textsuperscript{15}.

\subsubsection{Target estimands}\label{target-estimands}

Before describing our estimands for incident event reporting (in
paragraphs beginning with a bolded description below) we first describe
some key components. Within a group \(g\) and monitoring period \(m\),
\(F(t) = P(T \le t)\) is the cumulative distribution function of \(T\)
for all events beginning in that monitoring period. The overall hazard
\(\lambda (t) = P(T = t \mid T \ge t)\) can be defined in terms of
\(F(t)\) as \(\lambda (t) = (F(t) - F(t-1))/(1-F(t-1))\). In addition,
\(\overline{F}(t) = 1 - F(t) = P(T > t) = \prod_{t_i \le t} (1 - \lambda(t_i))\),
which is equivalent to the overall survival function in standard TTE
analyses\textsuperscript{20}.

The event-specific hazard is analogous to a cause-specific hazard in TTE
analyses with competing risks. For event \(s\) within group \(g\) and
monitoring period \(m\), the event-specific hazard at \(t_i\) is
\(\lambda_s (t_i) = P(T = t_i, S = s \mid T \ge t_i)\). Then, the
corresponding event-specific cumulative incidence is
\(F_s(t) = P(T \le t, S = s) = \sum_{t_i \le t} \overline{F}(t_i)\lambda_s(t_i) = \sum_{t_i \le t} \theta (t_i)\)\textsuperscript{20}.
Thus, the mean time without event \(s\) is
\(\mu_s(\tau) = \sum_{t_i < \tau} (t_{i+1} - t_i)F_s(t_i)\)\textsuperscript{29}.
If we are comparing across sequential monitoring periods or groups, we
write \(\mu_{s,m}(\tau)\) to specify the monitoring period \(m\) and
\(\mu_{s,g}(\tau)\) to specify group \(g\).

Right censoring is assumed to be non-informative except for when an
individual is censored due to a competing event. When an individual is
recorded as having an event \(s\) that has any competing events (e.g.,
if \(s > 1\)), they leave the risk set to be coded for any other
competing event. Both \(g=1\) and \(g=0\) groups are expected have
events recorded at all equivalent reporting times within the monitoring
period. We also impose that the overall sample is fixed across all
monitoring periods.

\textbf{Underreporting in comparison group.} The \(s^* \ge 1\) reference
events occur in \(g=0\) only. So, in the comparison group \(g=0\) only
and two sequential monitoring periods \(m-1\) and \(m\), our estimand
for the underreporting proportion \(\epsilon\) is one minus the average
persistence of all reference events, or
\(\epsilon = 1 - \frac{1}{h} \sum q_{s^*,m}\). Multiple pairs of
sequential monitoring periods are distinguished by adding a subscript,
\(\epsilon_m\), where \(m\) denotes the second monitoring period in a
pair. More details for this estimand can be found in the appendix.

\begin{figure}[H]

\centering{

\includegraphics[width=0.6\linewidth,height=\textheight,keepaspectratio]{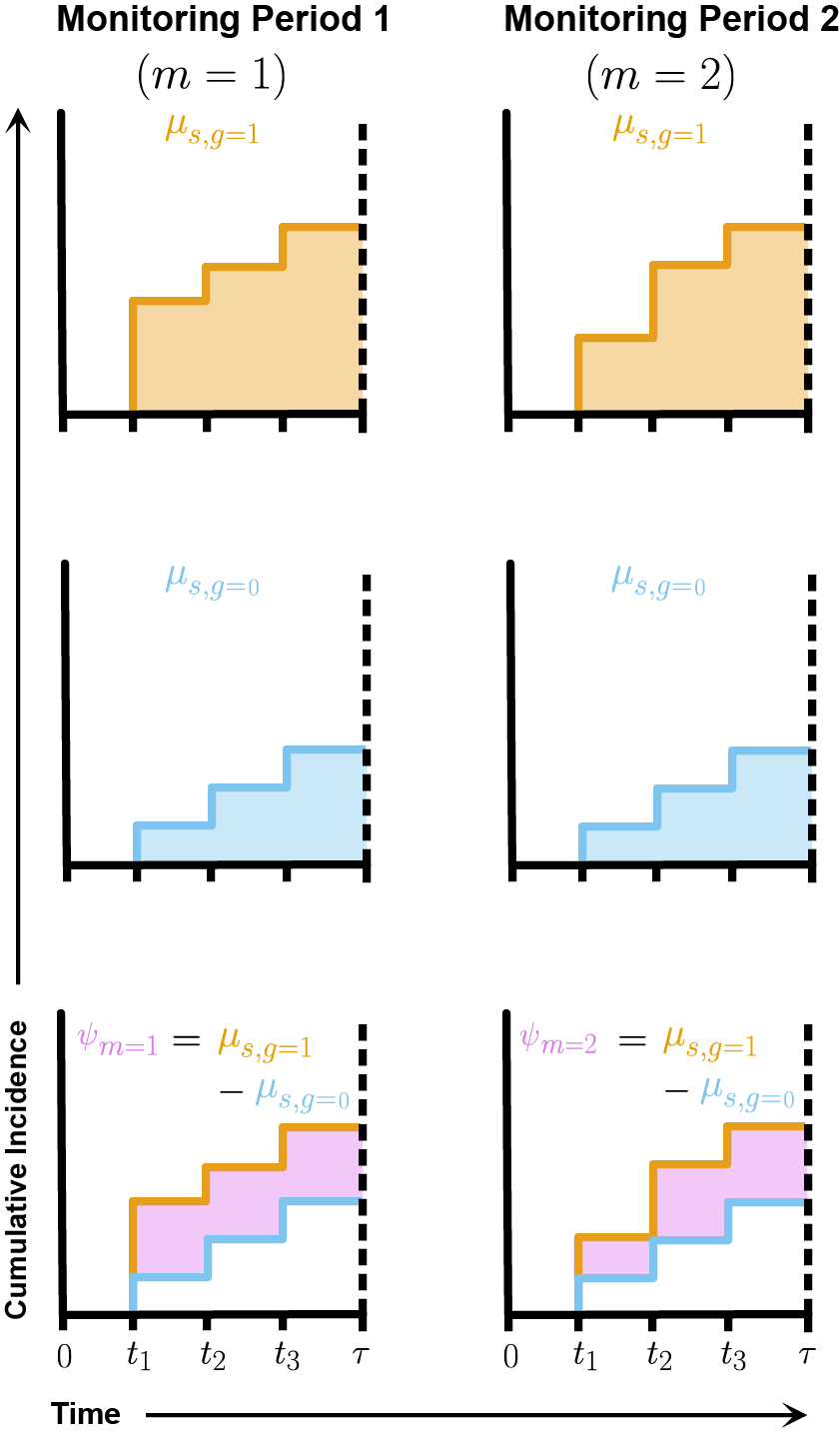}

}

\caption{\label{fig-schematic}\textbf{Schematic figure of notation for
evaluating coding intensity of the Hierarchical Condition Category (HCC)
denoted by} \(s\): This HCC is assumed to have no competing events. The
visuals and notation in this figure illustrate components of the
difference in mean time without event across groups estimator, or
\(\psi\), within two separate monitoring periods. \(\psi\) corresponds
to the difference of restricted mean time lost up until a pre-specified
end time \(\tau\) between a group \(g\) of interest (\(g=1\), e.g.,
Medicare Advantage) and a reference group (\(g=0\), e.g., Traditional
Medicare). Other reporting time points within each monitoring period are
denoted \(t_1\), \(t_2\), and \(t_3\) and also have an optional
subscript \(m\) to denote which monitoring period it corresponds to of
the two shown.}

\end{figure}

\textbf{Difference in mean time without event across groups.} We are
first interested in estimating the difference in incident reporting for
event \(s\) across the two groups within one monitoring period. So, our
estimand is \(\psi = \mu_{s,g=1}(\tau) - \mu_{s,g=0}(\tau)\). A
schematic illustration of this estimand as well as its underlying
components is available in Figure~\ref{fig-schematic}. In a monitoring
period where \(\epsilon\) is known, the estimand is modified to account
for underreporting by shifting the event-specific cumulative incidence
curve in the comparison group \(g=0\) by \(\epsilon\), or
\(\mu^*_{s,g=0}(\tau) = \sum_{t_i < \tau} (t_{i+1} - t_i)(F_s(t_i) + \epsilon)\).
This is written as \(\psi^* = \mu_{s,g=1}(\tau) - \mu^*_{s,g=0}(\tau)\).
Across the two groups in two sequential monitoring periods, the reported
difference in mean time without event that does not account for
underreporting is \(\psi_{M} = \psi_m - \psi_{m-1}\). \(\psi_m\) and
\(\psi_{m-1}\) are the same as \(\psi\) but with an additional subscript
to specify the monitoring period. This estimand can also be adjusted to
account for underreporting when \(\epsilon_m\) and \(\epsilon_{m-1}\)
are known, becoming \(\psi_M^* = \psi^*_m - \psi^*_{m-1}\).

\textbf{Possible severity-based upcoding.} Possible severity-based
upcoding can also be estimated within a monitoring period by comparing
the difference in incident reporting of the most severe event versus the
least severe event across groups. Here, \(s \ge 2\), where the least
severe subtype corresponds to \(s = 1\) and the most severe subtype to
\(s = k\). In one monitoring period, possible severity-based upcoding
across groups is \(\omega = \omega_{g=1} - \omega_{g=0}\), where each
\(\omega_{g} = \mu_{s=k,g}(\tau) - \mu_{s=1,g}(\tau)\). If such
reporting is higher in \(g=1\) compared with \(g=0\) (e.g.,
\(\omega > 0\)), then severity-based upcoding may be occurring.

\subsubsection{Estimators}\label{estimators}

In order to introduce the estimator of \(\mu_s(\tau)\), we first
describe the estimator for the event-specific hazard at \(t_i\):
\(\hat{\lambda}_s(t_i) = d_{s,i}/r_i\), where \(d_{s,i}\) is the count
of event \(s\) reported at \(t_i\) and \(r_i\) is the number of
individuals at risk at \(t_i\), meaning the number of individuals who
have not yet been recorded as having \(s\) or any competing event by
\(t_i\). \(d_i = \sum d_{s,i}\) is the count of all competing events
reported at \(t_i\). We also have the event-specific cumulative
incidence estimator:
\(\hat{F}_s(t) = \sum_{t_i \le t} \hat{\theta}(t_i)\), with
\(\hat{\theta}(t_i) = \widehat{\overline{F}}(t_i) \times \hat{\lambda}_s(t_i)\),
where
\(\widehat{\overline{F}}(t) = \prod_{t_i \le t}(1 - \hat{\lambda}(t_i)) = \prod_{t_i \le t}(1 - (d_i/r_i))\)
is the Kaplan-Meier estimator for
\(\overline{F}(t) = \prod_{t_i \le t}(1 - \lambda(t_i))\)\textsuperscript{20,37}.
We can now define the estimator of \(\mu_s(\tau)\) based on prior
literature\textsuperscript{29}, as this is a component of the novel
estimators we develop next:
\(\hat{\mu}_s(\tau) = \sum_{t_i < \tau} (t_{i+1} - t_i)\hat{F}_s(t_i)\).
Variances and covariances for estimators in this section are in the
appendix.

\textbf{Underreporting in comparison group.} For each reported reference
event \(s^*\) in \(g=0\) we first compute persistence, or the proportion
of individuals reported as having \(s^*\) in monitoring period \(m\) who
where also reported as having \(s^*\) in monitoring period \(m-1\). This
is then averaged across all reference events to obtain
\(\hat{\epsilon} = 1 - \frac{1}{h} \sum \hat{q}_{s^*,m}\).

\textbf{Difference in mean time without event across groups.} We propose
that the difference in mean time without an event, \(\psi\), is
estimated across groups as
\(\widehat{\psi} = \hat{\mu}_{s,g=1}(\tau) - \hat{\mu}_{s,g=0}(\tau)\).
To adjust for underreporting,
\(\hat{\mu}_{s,g=0}(\tau) = \sum_{t_i < \tau} (t_{i+1} - t_i) \hat{F}_{s}(t_i)\)
is modified to
\(\hat{\mu}_{s,g=0}(\tau) = \sum_{t_i < \tau} (t_{i+1} - t_i) (\hat{F}_{s}(t_i) + \epsilon)\)
in the estimator for \(\widehat{\psi}\); this is denoted
\(\hat{\psi}^*\). Across sequential monitoring periods, \(\psi_M\) is
estimated as \(\hat{\psi}_M = \hat{\psi}_m - \hat{\psi}_{m-1}\). When an
underreporting estimate is known for both monitoring periods, this
estimator becomes
\(\hat{\psi}_M^* = \hat{\psi}^*_m - \hat{\psi}^*_{m-1}\).

\textbf{Possible severity-based upcoding.} Our estimator \(\omega\), is
given by:
\(\widehat{\omega} = \widehat{\omega}_{g=1} - \widehat{\omega}_{g=0}\),
where each
\(\widehat{\omega}_{g} = \hat{\mu}_{s=k}(\tau) - \hat{\mu}_{s=1}(\tau)\)
within one monitoring period.

\subsection{Comparator estimator}\label{sec-comparator-estimator}

We compared our estimators to a coding intensity estimator similar to
the Demographic Estimate of Coding Intensity (DECI), which is widely
used by Medicare policymakers\textsuperscript{18,38}. DECI assumes
complete data (e.g., no censoring) and has the following formula:

\[
\begin{aligned}
\\ \text{DECI} = \frac{\frac{\text{National average MA CMS-HCC risk score}}{\text{National average TM CMS-HCC risk score}}}{\frac{\text{National average MA demographic-only CMS-HCC risk score}}{\text{National average TM demographic-only CMS-HCC risk score}}}.
\end{aligned}
\]

Here, the numerator is a risk score is estimated with both HCCs and
demographic variables, while the denominator is estimated uses
demographic variables only. Importantly, this also means that the
estimand targeted by DECI is different from that of our proposed
estimators. As we did not have CMS-HCC risk score coefficients or
demographic variables available in our simulated data, our comparator
estimator was more precisely DECI-like and defined as: \[
\begin{aligned}
\\ \text{DECI}^\dagger = \frac{\text{Average count of MA CMS-HCC HCCs}}{\text{Average count of TM CMS-HCC HCCs}}.
\end{aligned}
\]\(\text{DECI}^\dagger\) is estimated in the same simulated data
described earlier in this section, but relies on counts of all HCCs at
the end of each monitoring period rather than time to incident coding of
individual HCCs. Although DECI has a different target estimand than our
estimators and ignores censoring, comparing our simulation results to
\(\text{DECI}^\dagger\) is useful as it illustrates how our estimators
can be a complement to existing practice.

\subsection{\texorpdfstring{\texttt{upcoding} R
package}{upcoding R package}}\label{upcoding-r-package}

We developed the open source \texttt{upcoding} R package
(https://github.com/StanfordHPDS/upcoding) to help address several
issues in upcoding methods development by simulating longitudinal coding
data for a Medicare-eligible population. Features include:

\begin{enumerate}
\def\labelenumi{\arabic{enumi}.}
\item
  \textbf{Simulating a sample of individuals with realistic baseline
  HCCs}. To simulate realistic co-occurring HCCs not influenced by
  coding incentives, sets of self-reported health conditions from
  participants aged 65 years and older (i.e., Medicare eligible) were
  extracted from the National Institutes of Health's All of Us
  study\textsuperscript{39}. Baseline data were then simulated by
  sampling with replacement from these sets of co-occurring HCCs, where
  the respondent count weights sampling. The appendix includes survey
  summary statistics. This project has an Institutional Review Board
  exemption from All of Us because Stanford University has an
  institutional Data Use and Registration Agreement with the program,
  the data are anonymized, and we did not conduct human subjects
  research with the data. Our study also does not share any individual
  level All of Us data.
\item
  \textbf{Upcoding baseline data to a specified level over multiple time
  points.} Users have the option to upcode any HCC using either
  any-available or severity-based upcoding. This results in labeled
  upcoding data, which is a useful resource for many types of coding
  measurement and estimation.
\item
  \textbf{Undercoding baseline data to a specified level}. Users have
  the option to specify an undercoding proportion, and that proportion
  of all existing diagnoses are removed from the overall dataset. This
  can help simulate data that is similar to TM as a comparison group for
  analyses along with non-undercoded baseline data.
\end{enumerate}

We describe this functionality in further detail in the appendix.

\subsection{Simulation study}\label{sec-simulation-study}

Our simulation study demonstrates how the estimators we propose can be
used to monitor Medicare coding behaviors. To do this, different
upcoding and underreporting scenarios for the estimator of \(\psi\) were
compared to estimates of \(\text{DECI}^\dagger\). Degrees of upcoding
and undercoding were constructed to align with current estimates of
these issues in MA and TM from the literature. In addition, the temporal
structure was implemented to be analogous to quarterly reporting for the
length of time (around two years) that a given risk adjustment formula
version is typically in place. We did not use individual-level Medicare
data given that labeled upcoding is not available. Each individual
scenario was replicated 1000 times.

In each replicate, two sets of 1,000,000 observations of baseline data
were independently simulated using our \texttt{upcoding} package. For
each dataset, the columns are the V28 HCCs (listed in appendix). Both
any-available and severity-based upcoding were implemented in the first
baseline data set (the MA data). For the former type of upcoding, an HCC
without any competing HCCs (HCC238, Specified Heart Arrhythmias) was
upcoded. For the latter we upcoded HCC125 (Dementia, Severe). HCC125's
less severe HCCs are HCC126 (Dementia, Moderate) and HCC127 (Dementia,
Mild or Unspecified). The scenarios implemented were as follows:

\textbf{Scenario 1. Any-available upcoding of an HCC that lacks
competing events (HCC238) with varying underreporting in the comparison
group.} Baseline MA data were separately upcoded to varying degrees
(20\%, 25\%, 30\%) sequentially within each monitoring period. The
comparison group (i.e., TM data) was simulated by first undercoding the
baseline TM data to varying levels (0\%, 5\%, 10\%, 15\%) and then
upcoding HCC238 analogously but to a lower amount (5\%) per monitoring
period.

\textbf{Scenario 2. Lower-severity upcoding of an HCC with competing
events (HCC125) with varying underreporting in the comparison group.}
Baseline MA data were separately upcoded to varying degrees (20\%, 25\%,
30\%) sequentially within each monitoring period. The comparison group
was simulated by first undercoding the baseline TM data to varying
levels (0\%, 5\%, 10\%, 15\%) and then upcoding HCC125 analogously but
to a lower amount (5\%) per monitoring period.

\section{Results}\label{results}

\subsection{Proposed estimators}\label{proposed-estimators}

\begin{figure}[H]

\centering{

\includegraphics[width=0.9\linewidth,height=\textheight,keepaspectratio]{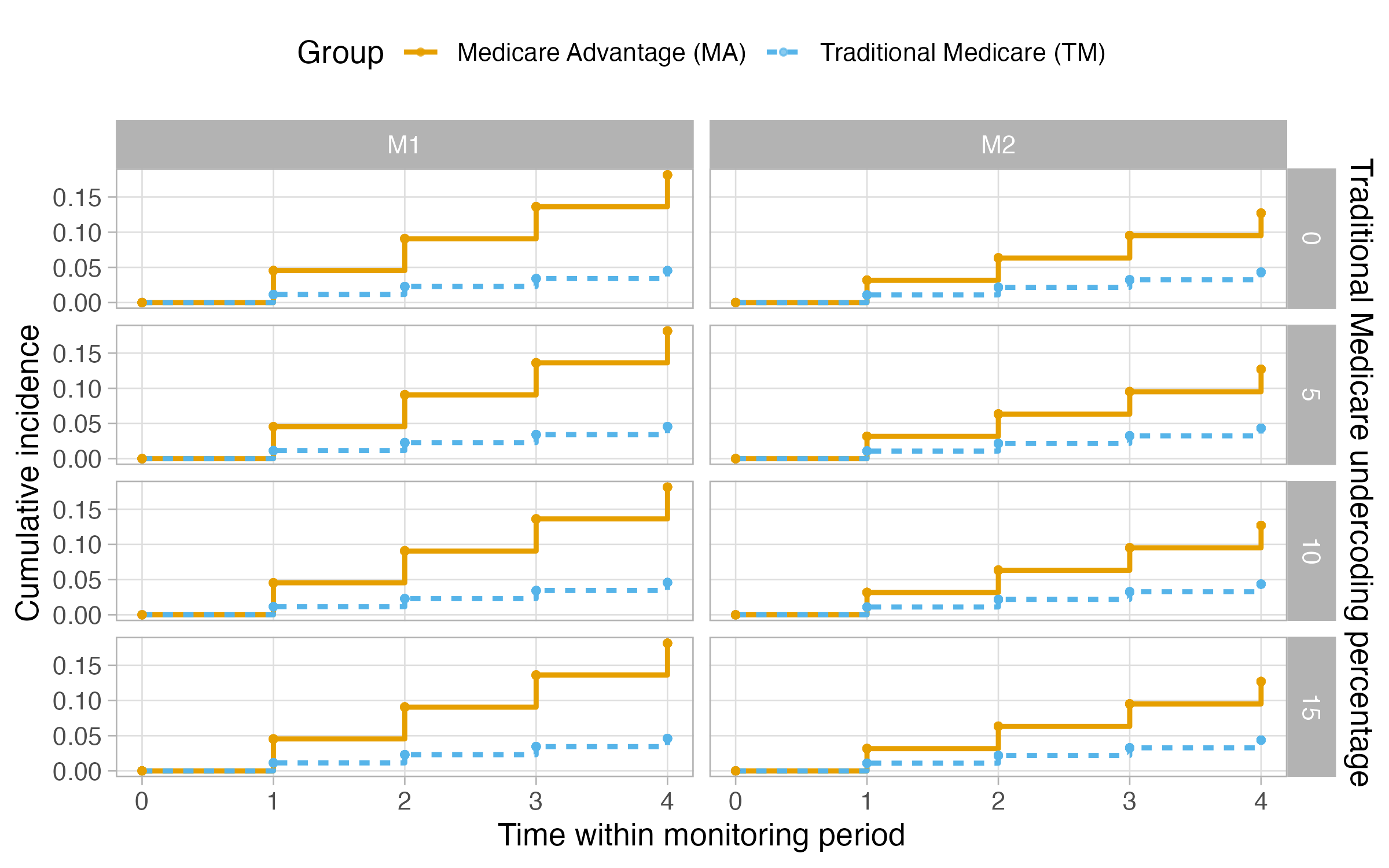}

}

\caption{\label{fig-cif}\textbf{Cumulative incidence functions for the
Specified Heart Arrhythmias Hierarchical Condition Category (HCC) in
simulated Medicare Advantage (MA) and Traditional Medicare (TM) groups:
20\% any-available MA upcoding.} 20\% any-available MA upcoding.
Specified heart arrhythmias corresponds to HCC238, which does not have
any competing events. For this HCC, 20\% of any-available individuals in
the MA group are upcoded and 5\% of any-available individuals in the TM
comparison group are upcoded. The first monitoring period is labeled M1,
and the second monitoring period is labeled M2. Given the large sample
size, confidence intervals are very narrow and are therefore omitted as
they cannot be distinguished visually.}

\end{figure}

Cumulative incidence of coding for scenario 1 given 20\% any-available
upcoding in MA and 5\% any-available upcoding in TM after all four
degrees of undercoding is shown in Figure~\ref{fig-cif}. As expected,
within each monitoring period the cumulative incidence of the upcoded
group was higher than the comparison group, which was only upcoded 5\%.
We also saw that the impact of undercoding on cumulative incidence
estimates was limited. Given that undercoding occurs across the entire
set of 115 HCCs, it was less likely to notably influence estimates for
any single HCC. Lastly, we observed that the gaps between upcoded and
comparison groups become smaller in sequential monitoring periods, which
is a consequence of the fixed sample we imposed across all monitoring
periods. As monitoring periods increased, the number of individuals
available to upcode decreased. Results plots for additional degrees of
upcoding (25\%, 30\%) and Scenario 2 upcoding (severity-based upcoding
for HCC125) are in the appendix.

\begin{figure}[H]

\centering{

\includegraphics[width=0.9\linewidth,height=\textheight,keepaspectratio]{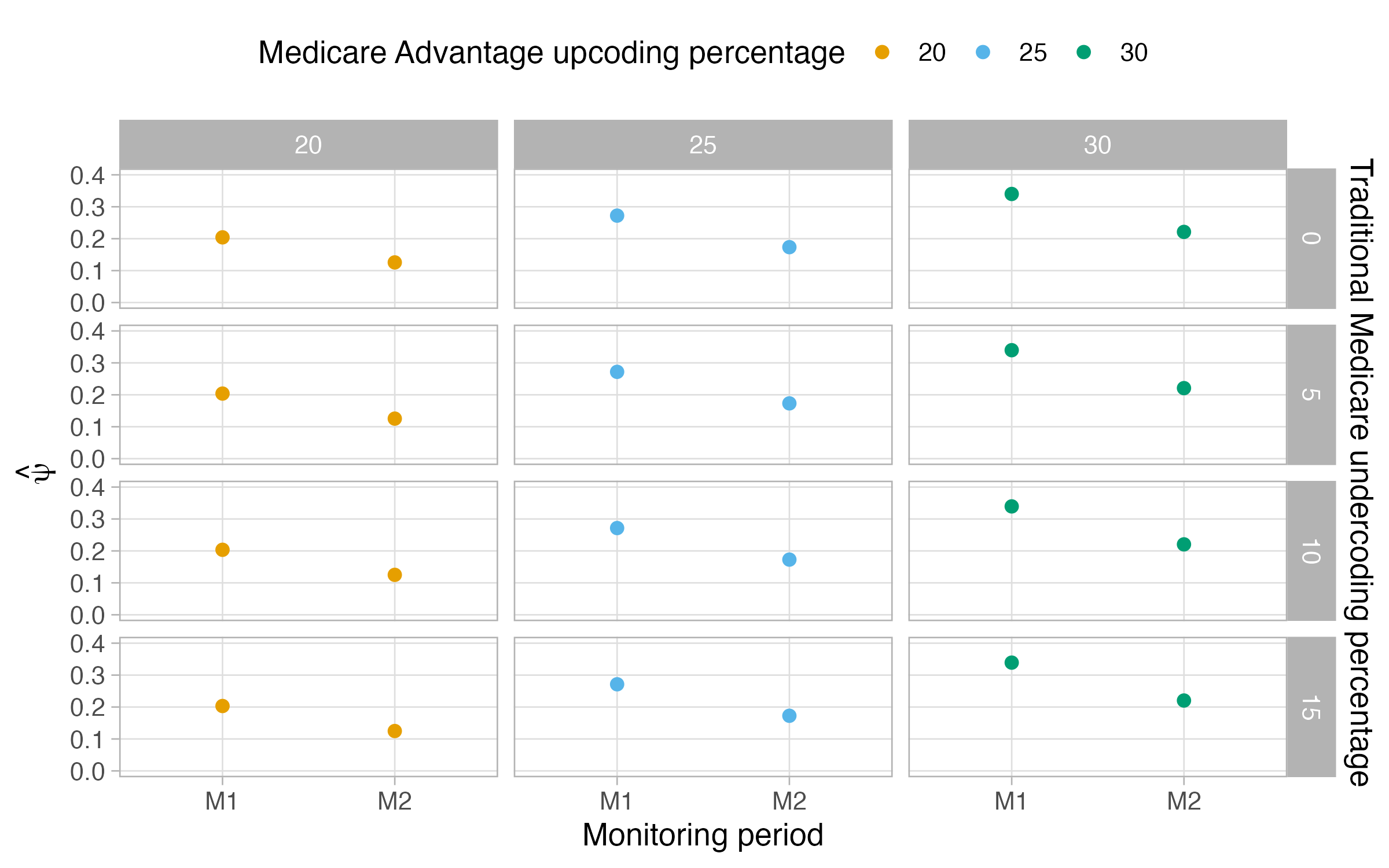}

}

\caption{\label{fig-psi}\textbf{Within-monitoring period period}
\(\boldsymbol{\psi}\) \textbf{estimates} \textbf{for the Specified Heart
Arrhythmias Hierarchical Condition Category (HCC) in simulated Medicare
Advantage (MA) and Traditional Medicare (TM) groups.} \(\widehat{\psi}\)
corresponds to the proposed estimator for \(\psi\), or the difference in
mean time without event (i.e., incident coding of the HCC) across
groups. Specified heart arrhythmias corresponds to HCC238, which does
not have any competing events. For this HCC, any-available individuals
in the MA group are upcoded to varying degrees and any-available
individuals in the TM comparison group are upcoded 5\%. The first
monitoring period is labeled M1, and the second monitoring period is
labeled M2. Given the large sample size, confidence intervals are very
narrow and are therefore omitted as they cannot be distinguished
visually.}

\end{figure}

Estimates corresponding to the \(\psi\) estimator for upcoding of HCC238
are presented in Figure~\ref{fig-psi}, which examines the difference in
time without reported incident HCC238 coding between MA and TM within
each monitoring period. Examining the first monitoring period (M1),
estimates clearly increase with the degree of upcoding. Similarly, for
monitoring period 2 (M2) alone, estimates also increase with the degree
of upcoding.

In addition, undercoding has a limited effect on estimates. However,
researchers also have the option to adjust for underreporting using the
estimator proposed for underreporting. As the sample is fixed across
monitoring periods, estimates across monitoring periods within a degree
of upcoding and undercoding decrease, analogously to
Figure~\ref{fig-cif}. The gap between M1 and M2 increases with degree of
upcoding as well, because more aggressive incident upcoding in M1 limits
the availability of individuals for incident upcoding in M2.

This suggests that within a monitoring period, researchers could compare
estimates of different HCCs to obtain a ranked list of HCCs to study
further, where HCCs with the highest estimates have the most incident
coding. In addition, in this scenario, being in TM appeared to have a
``protective effect'' against being coded with HCC238, although
supplemental analyses would be needed to account for possible
confounding and other issues. Additional results plots for
severity-based upcoding of HCC125 are available in the appendix.

\subsection{Comparator estimator}\label{comparator-estimator}

\begin{figure}[H]

\centering{

\includegraphics[width=0.9\linewidth,height=\textheight,keepaspectratio]{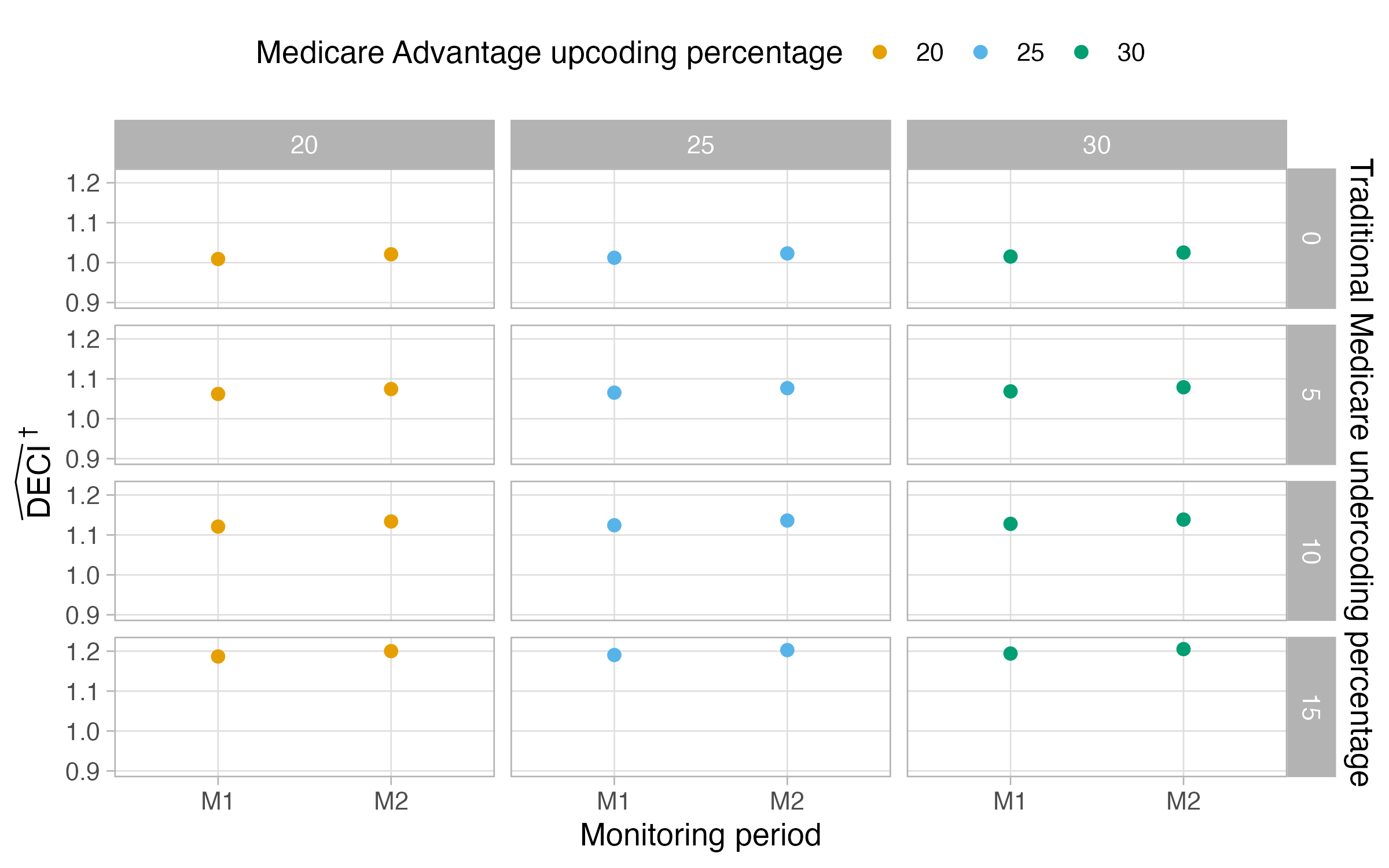}

}

\caption{\label{fig-deci}\(\textbf{DECI}^\dagger\) \textbf{estimate
across all Medicare Advantage (MA) Version 28 Hierarchical Condition
Categories (HCCs) at varying degrees of upcoding and underreporting in
simulated MA and Traditional Medicare (TM) groups.} The Demographic
Estimate of Coding Intensity (DECI) is an estimator widely used by
policymakers to evaluate coding intensity within a year by comparing
ratios of average risk scores in MA versus TM; \(\text{DECI}^\dagger\)
is our slight modification of this estimand due to data availability,
and \(\widehat{\text{DECI}}^\dagger\) is its estimate. Three separate
degrees of MA group upcoding occur sequentially over each monitoring
period in HCC238 (any-available) and HCC125 (lower severity) only. The
TM group is upcoded 5\% sequentially for the same HCCs over equivalent
periods. The first monitoring period is labeled M1, and the second
monitoring period is labeled M2. Given the large sample size, confidence
intervals are very narrow and are therefore omitted as they cannot be
distinguished visually.}

\end{figure}

Figure~\ref{fig-deci} shows the average \(\text{DECI}^\dagger\) estimate
across simulation replicates by monitoring period as well as upcoding
and undercoding degree. As degree of upcoding increases, these estimates
increase negligibly both for individual monitoring periods and across
sequential monitoring periods. This is intuitive--only two of the 115
HCCs in the current MA risk adjustment algorithm are being upcoded--but
suggests a limitation of DECI in identifying upcoding occurring in a
minority of HCCs. Regardless of upcoding degree, the
\(\text{DECI}^\dagger\) estimate increasing in line with the degree of
TM undercoding indicates that it may be more sensitive to undercoding in
TM. Especially since we know undercoding is prevalent in
TM\textsuperscript{3,16}, this suggests that DECI may be underestimating
overall spending gap between MA and TM.

\section{Discussion}\label{discussion}

We proposed a set of estimators that extend RMTL methodology for
evaluating time to incident coding by private insurers in Medicare.
Given the timeline of CMS-HCC updates, our approach realistically
presupposes that reported coding is examined over multiple monitoring
periods, each of which has several time points where new events are
reported. Novel estimators were introduced to evaluate differences in
time-to-reporting both within monitoring periods and across monitoring
periods, including an estimator for severity-based upcoding. Our
approach also included an adjustment for possible underreporting, which
is a known major issue in TM data\textsuperscript{3,16}. Finally, we
developed an open source R package that enables users to simulate
co-occurring HCCs free of coding incentives and similar to those
reported by individuals residing in the United States eligible for
Medicare. Users can also undercode or upcode this data over time.

Our simulated data were upcoded to degrees aligned to those reported in
literature\textsuperscript{7}, and we found in our simulation results
that our estimators were able to recover differences in upcoding both
within and across monitoring periods while showing limited sensitivity
to undercoding. DECI-like estimates of the same data were a useful
complement, but were limited in that these comparator estimates were
very sensitive to undercoding and did not vary when upcoding occurred in
a minority of HCCs. Therefore, our estimators show considerable promise
as tools to help evaluate reported coding patterns over time. These
estimators can be used as a first step to identify potentially upcoded
HCCs while making more realistic assumptions than DECI. Follow up
analyses could include examining coding patterns within specific
insurers, providers, or beneficiaries.

This work has a number of limitations and areas for future development.
First, several assumptions could be further relaxed. This includes our
assumptions that the population is fixed across both comparison groups
and monitoring periods and that estimates in sequential monitoring
periods across groups are approximately independent. However, doing so
could also make establishing a diagnosis as incident and comparing
estimates across monitoring periods more challenging. Upcoding
estimators could be expanded to additional types of upcoding and to
correct for underreporting. Covariates could also be incorporated into
estimation to better address issues like confounding. There is also more
heterogeneity and missingness in real-world claims data than was
included in our simulations. Self-reported diagnoses---as we use in our
simulated baseline data---also have recall bias and other issues that we
do not address here.

Since the MA program's inception, high intensity of private insurer
coding, including possible upcoding, is estimated to have cost the
federal government and taxpayers \$224 billion dollars without clear
benefit to beneficiaries\textsuperscript{7}. Thus, estimating
differences in coding between MA and TM or more and less severe
competing HCCs has the potential to help policymakers locate issues with
specific HCCs earlier and at scale. This could suggest areas for
improvement in the risk adjustment formula and potentially save
significant funds in the Medicare program. Our work has provided both
novel estimators and novel simulated data tailored to addressing these
important policy considerations.

\newpage{}

\section{Figures}\label{figures}

\newpage{}

\newpage{}

\newpage{}

\newpage{}

\section{Acknowledgments}\label{acknowledgments}

This research was funded by National Institutes of Health Director's
Pioneer Award DP1LM014278 and a Stanford Interdisciplinary Graduate
Fellowship. We thank Malcolm Barrett, Gabriela Basel, Lizzie Kumar, and
Neha Srivathsa for their valuable insights and contributions to code
performance and review. We gratefully acknowledge All of Us participants
for their contributions, without whom this research would not have been
possible. We also thank the National Institutes of Health's All of Us
Research Program (https://allofus.nih.gov/) for making available the
participant survey data included in this study. We thank Stanford
University and Stanford Research Computing for providing additional
computational resources, including support for computing performed on
the Sherlock cluster.

\section{Conflict of interest
statement}\label{conflict-of-interest-statement}

None to report.

\newpage{}

\section*{References}\label{references}
\addcontentsline{toc}{section}{References}

\phantomsection\label{refs}


\begin{CSLReferences}{0}{1}
\bibitem[\citeproctext]{ref-Ochieng2025-bv}
\CSLLeftMargin{1. }%
\CSLRightInline{Ochieng N, Cubanski J, Neuman T. {A Snapshot of Sources
of Coverage Among Medicare Beneficiaries}. Published online December
2025. Accessed January 1, 2025.
\url{https://www.kff.org/medicare/issue-brief/a-snapshot-of-sources-of-coverage-among-medicare-beneficiaries/}}

\bibitem[\citeproctext]{ref-Pope2004-tv}
\CSLLeftMargin{2. }%
\CSLRightInline{Pope GC, Kautter J, Ellis RP, et al. {Risk Adjustment of
Medicare Capitation Payments Using the CMS-HCC Model}. \emph{Health Care
Financ Rev}. 2004;25(4):119-141.
\url{https://pmc.ncbi.nlm.nih.gov/articles/PMC4194896/pdf/hcfr-25-4-119.pdf}}

\bibitem[\citeproctext]{ref-Ellis2018-mr}
\CSLLeftMargin{3. }%
\CSLRightInline{Ellis RP, Martins B, Rose S. {Chapter 3 - Risk
Adjustment for Health Plan Payment}. In: McGuire TG, Kleef RC van, eds.
\emph{{Risk Adjustment, Risk Sharing and Premium Regulation in Health
Insurance Markets}}. Academic Press; 2018:55-104.
doi:\href{https://doi.org/10.1016/B978-0-12-811325-7.00003-8}{10.1016/B978-0-12-811325-7.00003-8}}

\bibitem[\citeproctext]{ref-Centers_for_Medicare_and_Medicaid_Services2023-zf}
\CSLLeftMargin{4. }%
\CSLRightInline{Centers for Medicare \& Medicaid Services. {Advance
Notice of Methodological Changes for Calendar Year (CY) 2024 for
Medicare Advantage (MA) Capitation Rates and Part C and Part D Payment
Policies}. Published online February 1, 2023. Accessed June 15, 2024.
\url{https://www.cms.gov/files/document/2024-advance-notice-pdf.pdf}}

\bibitem[\citeproctext]{ref-geruso2020}
\CSLLeftMargin{5. }%
\CSLRightInline{Geruso M, Layton T. {Upcoding: Evidence from Medicare on
Squishy Risk Adjustment}. \emph{{J Polit Econ}}. 2020;128(3):984-1026.
doi:\href{https://doi.org/10.1086/704756}{10.1086/704756}}

\bibitem[\citeproctext]{ref-joiner2024}
\CSLLeftMargin{6. }%
\CSLRightInline{Joiner KA, Lin J, Pantano J. {Upcoding in Medicare:
Where Does it Matter Most?} \emph{{Health Econ Rev}}. 2024;14(1).
doi:\href{https://doi.org/10.1186/s13561-023-00465-4}{10.1186/s13561-023-00465-4}}

\bibitem[\citeproctext]{ref-Medpac2025-uh}
\CSLLeftMargin{7. }%
\CSLRightInline{Medicare Payment Advisory Commission. {Report to the
Congress: Medicare Payment Policy}. Published online March 2025.
Accessed March 17, 2025.
\url{https://www.medpac.gov/wp-content/uploads/2025/03/Mar25_MedPAC_Report_To_Congress_SEC-1.pdf}}

\bibitem[\citeproctext]{ref-Abelson2023-gr}
\CSLLeftMargin{8. }%
\CSLRightInline{Abelson R, Sanger-Katz M. {New Medicare Rule Aims to
Take Back \$4.7 Billion from Insurers}. \emph{The New York Times}.
Published online January 2023. Accessed February 1, 2023.
\url{https://www.nytimes.com/2023/01/30/upshot/medicare-overbilling-biden-rule.html}}

\bibitem[\citeproctext]{ref-Centers_for_Medicare_Medicaid_Services2023-dy}
\CSLLeftMargin{9. }%
\CSLRightInline{Centers for Medicare \& Medicaid Services, United States
Department of Health and Human Services. {Medicare and Medicaid
Programs; Policy and Technical Changes to the Medicare Advantage,
Medicare Prescription Drug Benefit, Program of All-Inclusive Care for
the Elderly (PACE), Medicaid Fee-For-Service, and Medicaid Managed Care
Programs for Years 2020 and 2021}. \emph{Federal Register 88 FR 6643}.
Published online February 1, 2023:6643-6665. Accessed June 5, 2023.
\url{https://www.federalregister.gov/d/2023-01942}}

\bibitem[\citeproctext]{ref-Constantino2025-qf}
\CSLLeftMargin{10. }%
\CSLRightInline{Constantino AK. {UnitedHealth Says it is Cooperating
with DOJ Investigations into Medicare Billing Practices}. \emph{Consumer
News and Business Channel}. Published online July 24, 2025. Accessed
November 6, 2025.
\url{https://www.cnbc.com/2025/07/24/unitedhealthcare-doj-investigation-medicare-billing.html}}

\bibitem[\citeproctext]{ref-DOJ2026-vd}
\CSLLeftMargin{11. }%
\CSLRightInline{United States Department of Justice, Office of Public
Affairs. {Kaiser Permanente Affiliates Pay \$556M to Resolve False
Claims Act Allegations}. Published online January 14, 2026. Accessed
January 15, 2026.
\url{https://www.justice.gov/opa/pr/kaiser-permanente-affiliates-pay-556m-resolve-false-claims-act-allegations}}

\bibitem[\citeproctext]{ref-rose2016}
\CSLLeftMargin{12. }%
\CSLRightInline{Rose S, Zaslavsky AM, McWilliams JM. Variation in
Accountable Care Organization Spending and Sensitivity to Risk
Adjustment: Implications for Benchmarking. \emph{Health Aff}.
2016;35(3):440-448.
doi:\href{https://doi.org/10.1377/hlthaff.2015.1026}{10.1377/hlthaff.2015.1026}}

\bibitem[\citeproctext]{ref-chernew2021}
\CSLLeftMargin{13. }%
\CSLRightInline{Chernew ME, Carichner J, Impreso J, et al. Coding-Driven
Changes in Measured Risk in Accountable Care Organizations. \emph{Health
Aff}. 2021;40(12):1909-1917.
doi:\href{https://doi.org/10.1377/hlthaff.2021.00361}{10.1377/hlthaff.2021.00361}}

\bibitem[\citeproctext]{ref-McWilliams2025-hz}
\CSLLeftMargin{14. }%
\CSLRightInline{McWilliams JM, Weinreb G, Landrum MB, Chernew ME. {Use
of Patient Health Survey Data for Risk Adjustment to Limit Distortionary
Coding Incentives in Medicare}. \emph{Health Aff}. 2025;44(1):48-57.
doi:\href{https://doi.org/10.1377/hlthaff.2023.01351}{10.1377/hlthaff.2023.01351}}

\bibitem[\citeproctext]{ref-McGuire2025-bp}
\CSLLeftMargin{15. }%
\CSLRightInline{McGuire TG, Enache OM, Chernew M, McWilliams JM, Nham T,
Rose S. {Incidence, Persistence, and Steady-State Prevalence in Coding
Intensity for Health Plan Payment}. \emph{Health Serv Res}.
2026;61(1):e70065.
doi:\href{https://doi.org/10.1111/1475-6773.70065}{10.1111/1475-6773.70065}}

\bibitem[\citeproctext]{ref-Ghoshal-Datta2024-io}
\CSLLeftMargin{16. }%
\CSLRightInline{Ghoshal-Datta N, Chernew ME, McWilliams JM. {Lack of
Persistent Coding in Traditional Medicare May Widen the Risk-Score Gap
with Medicare Advantage}. \emph{Health Aff}. 2024;43(12):1638-1646.
doi:\href{https://doi.org/10.1377/hlthaff.2024.00169}{10.1377/hlthaff.2024.00169}}

\bibitem[\citeproctext]{ref-jacobs2018}
\CSLLeftMargin{17. }%
\CSLRightInline{Jacobs PD, Kronick R. {Getting What We Pay For: How Do
Risk-Based Payments to Medicare Advantage Plans Compare with Alternative
Measures of Beneficiary Health Risk?} \emph{Health Serv Res}.
2018;53(6):4997-5015.
doi:\href{https://doi.org/10.1111/1475-6773.12977}{10.1111/1475-6773.12977}}

\bibitem[\citeproctext]{ref-kronick2021a}
\CSLLeftMargin{18. }%
\CSLRightInline{Kronick R, Chua FM. Industry-Wide and Sponsor-Specific
Estimates of Medicare Advantage Coding Intensity. \emph{SSRN Electronic
Journal}. Published online 2021.
doi:\href{https://doi.org/10.2139/ssrn.3959446}{10.2139/ssrn.3959446}}

\bibitem[\citeproctext]{ref-James2021-ej}
\CSLLeftMargin{19. }%
\CSLRightInline{James G, Witten D, Hastie T, Tibshirani R. \emph{{An
Introduction to Statistical Learning: with Applications in R}}. 2nd ed.
Springer; 2021.}

\bibitem[\citeproctext]{ref-Geskus2020-tb}
\CSLLeftMargin{20. }%
\CSLRightInline{Geskus RB. \emph{Data Analysis with Competing Risks and
Intermediate States}. 1st ed. Chapman \& Hall/CRC; 2015.}

\bibitem[\citeproctext]{ref-Hernan2010-qs}
\CSLLeftMargin{21. }%
\CSLRightInline{Hernán MA. {The Hazards of Hazard Ratios}.
\emph{Epidemiology}. 2010;21(1):13-15.
doi:\href{https://doi.org/10.1097/EDE.0b013e3181c1ea43}{10.1097/EDE.0b013e3181c1ea43}}

\bibitem[\citeproctext]{ref-Chappell2016-gr}
\CSLLeftMargin{22. }%
\CSLRightInline{Chappell R, Zhu X. {Describing Differences in Survival
Curves}. \emph{JAMA Oncol}. 2016;2(7):906-907.
doi:\href{https://doi.org/10.1001/jamaoncol.2016.0001}{10.1001/jamaoncol.2016.0001}}

\bibitem[\citeproctext]{ref-Uno2020-hk}
\CSLLeftMargin{23. }%
\CSLRightInline{Uno H, Horiguchi M, Hassett MJ. {Statistical
Test/Estimation Methods Used in Contemporary Phase III Cancer Randomized
Controlled Trials with Time-to-Event Outcomes}. \emph{Oncologist}.
2020;25(2):91-93.
doi:\href{https://doi.org/10.1634/theoncologist.2019-0464}{10.1634/theoncologist.2019-0464}}

\bibitem[\citeproctext]{ref-Jachno2019-bk}
\CSLLeftMargin{24. }%
\CSLRightInline{Jachno K, Heritier S, Wolfe R. {Are Non-Constant Rates
and Non-Proportional Treatment Effects Accounted For in the Design and
Analysis of Randomised Controlled Trials? A Review of Current Practice}.
\emph{{BMC Med Res Methodol}}. 2019;19(1):103.
doi:\href{https://doi.org/10.1186/s12874-019-0749-1}{10.1186/s12874-019-0749-1}}

\bibitem[\citeproctext]{ref-Tian2014-xr}
\CSLLeftMargin{25. }%
\CSLRightInline{Tian L, Zhao L, Wei LJ. {Predicting the Restricted Mean
Event Time with the Subject's Baseline Covariates in Survival Analysis}.
\emph{Biostatistics}. 2014;15(2):222-233.
doi:\href{https://doi.org/10.1093/biostatistics/kxt050}{10.1093/biostatistics/kxt050}}

\bibitem[\citeproctext]{ref-Trinquart2016-cv}
\CSLLeftMargin{26. }%
\CSLRightInline{Trinquart L, Jacot J, Conner SC, Porcher R. {Comparison
of Treatment Effects Measured by the Hazard Ratio and by the Ratio of
Restricted Mean Survival Times in Oncology Randomized Controlled
Trials}. \emph{{J Clin Oncol}}. 2016;34(15):1813-1819.
doi:\href{https://doi.org/10.1200/JCO.2015.64.2488}{10.1200/JCO.2015.64.2488}}

\bibitem[\citeproctext]{ref-Tian2020-sw}
\CSLLeftMargin{27. }%
\CSLRightInline{Tian L, Jin H, Uno H, et al. {On the Empirical Choice of
the Time Window for Restricted Mean Survival Time}. \emph{Biometrics}.
2020;76(4):1157-1166.
doi:\href{https://doi.org/10.1111/biom.13237}{10.1111/biom.13237}}

\bibitem[\citeproctext]{ref-Austin2017-nm}
\CSLLeftMargin{28. }%
\CSLRightInline{Austin PC, Fine JP. {Accounting for Competing Risks in
Randomized Controlled Trials: a Review and Recommendations for
Improvement}. \emph{Stat Med}. 2017;36(8):1203-1209.
doi:\href{https://doi.org/10.1002/sim.7215}{10.1002/sim.7215}}

\bibitem[\citeproctext]{ref-Conner2021-ac}
\CSLLeftMargin{29. }%
\CSLRightInline{Conner SC, Trinquart L. {Estimation and Modeling of the
Restricted Mean Time Lost in the Presence of Competing Risks}.
\emph{Stat Med}. 2021;40(9):2177-2196.
doi:\href{https://doi.org/10.1002/sim.8896}{10.1002/sim.8896}}

\bibitem[\citeproctext]{ref-Irwin1949-ez}
\CSLLeftMargin{30. }%
\CSLRightInline{Irwin JO. {The Standard Error of an Estimate of
Expectation of Life, with Special Reference to Expectation of Tumourless
Life in Experiments with Mice}. \emph{{J Hyg (Lond)}}. 1949;47(2):188.
doi:\href{https://doi.org/10.1017/s0022172400014443}{10.1017/s0022172400014443}}

\bibitem[\citeproctext]{ref-Royston2011-sz}
\CSLLeftMargin{31. }%
\CSLRightInline{Royston P, Parmar MKB. {The Use of Restricted Mean
Survival Time to Estimate the Treatment Effect in Randomized Clinical
Trials when the Proportional Hazards Assumption is in Doubt}. \emph{Stat
Med}. 2011;30(19):2409-2421.
doi:\href{https://doi.org/10.1002/sim.4274}{10.1002/sim.4274}}

\bibitem[\citeproctext]{ref-Uno2014-ul}
\CSLLeftMargin{32. }%
\CSLRightInline{Uno H, Claggett B, Tian L, et al. {Moving Beyond the
Hazard Ratio in Quantifying the Between-Group Difference in Survival
Analysis}. \emph{J Clin Oncol}. 2014;32(22):2380-2385.
doi:\href{https://doi.org/10.1200/JCO.2014.55.2208}{10.1200/JCO.2014.55.2208}}

\bibitem[\citeproctext]{ref-Zhao2016-ca}
\CSLLeftMargin{33. }%
\CSLRightInline{Zhao L, Claggett B, Tian L, et al. {On the Restricted
Mean Survival Time Curve in Survival Analysis}. \emph{Biometrics}.
2016;72(1):215-221.
doi:\href{https://doi.org/10.1111/biom.12384}{10.1111/biom.12384}}

\bibitem[\citeproctext]{ref-Andersen2013-jm}
\CSLLeftMargin{34. }%
\CSLRightInline{Andersen PK. {Decomposition of Number of Life Years Lost
According to Causes of Death}. \emph{Stat Med}. 2013;32(30):5278-5285.
doi:\href{https://doi.org/10.1002/sim.5903}{10.1002/sim.5903}}

\bibitem[\citeproctext]{ref-Calkins2018-ed}
\CSLLeftMargin{35. }%
\CSLRightInline{Calkins KL, Canan CE, Moore RD, Lesko CR, Lau B. {An
Application of Restricted Mean Survival Time in a Competing Risks
Setting: Comparing Time to ART Initiation by Injection Drug Use}.
\emph{BMC Med Res Methodol}. 2018;18(1):27.
doi:\href{https://doi.org/10.1186/s12874-018-0484-z}{10.1186/s12874-018-0484-z}}

\bibitem[\citeproctext]{ref-Wu2022-mx}
\CSLLeftMargin{36. }%
\CSLRightInline{Wu H, Yuan H, Yang Z, Hou Y, Chen Z. {Implementation of
an Alternative method for Assessing Competing Risks: Restricted Mean
Time Lost}. \emph{Am J Epidemiol}. 2022;191(1):163-172.
doi:\href{https://doi.org/10.1093/aje/kwab235}{10.1093/aje/kwab235}}

\bibitem[\citeproctext]{ref-Kaplan1958-ye}
\CSLLeftMargin{37. }%
\CSLRightInline{Kaplan EL, Meier P. {Nonparametric Estimation from
Incomplete Observations}. \emph{J Am Stat Assoc}. 1958;53(282):457-481.
doi:\href{https://doi.org/10.1080/01621459.1958.10501452}{10.1080/01621459.1958.10501452}}

\bibitem[\citeproctext]{ref-Medicare_Payment_Advisory_Commission2024-cg}
\CSLLeftMargin{38. }%
\CSLRightInline{Medicare Payment Advisory Commission. {Report to the
Congress: Medicare Payment Policy}. Published online March 24, 2024.
Accessed May 1, 2024.
\url{https://www.medpac.gov/wp-content/uploads/2024/03/Mar24_MedPAC_Report_To_Congress_SEC.pdf}}

\bibitem[\citeproctext]{ref-theall2019}
\CSLLeftMargin{39. }%
\CSLRightInline{All of Us Research Program Investigators. {The {``All of
Us''} Research Program}. \emph{N Engl J Med}. 2019;381(7):668-676.
doi:\href{https://doi.org/10.1056/NEJMsr1809937}{10.1056/NEJMsr1809937}}

\end{CSLReferences}
\end{document}


\begin{frontmatter}
\title{Appendix \\\large{Time-to-Event Estimation with Unreliably
Reported Events in Medicare Health Plan Payment} }
\author[1]{Oana M. Enache%
\corref{cor1}%
}
 \ead{oenache@stanford.edu} 
\author[2]{Sherri Rose%
}

\affiliation[1]{organization={Stanford University School of
Medicine, Department of Biomedical Data Science},addressline={Edwards
Building, 300 Pasteur Drive},city={Stanford,
CA},postcode={94304},postcodesep={}}
\affiliation[2]{organization={Stanford University, Department of Health
Policy},addressline={Encina Commons, 615 Crothers Way},city={Stanford,
CA},postcode={94305},postcodesep={}}

\cortext[cor1]{Corresponding author}

\end{frontmatter}

\setstretch{1}
\section{Considerations for underreporting
estimand}\label{considerations-for-underreporting-estimand}

The underreporting estimand is used to uniformly shift cumulative
incidence in the comparison group (\(g=0\), e.g.~Traditional Medicare,
TM) only. Aligned with prior work\textsuperscript{1,2}, we recommend
choosing reference events for underreporting estimation that are
expected to persist over time (e.g.~paraplegia, human immunodeficiency
virus/acquired immunodeficiency syndrome). However, because reference
events should be distinct from Hierarchical Condition Categories (HCCs)
used for the other forms of estimation we propose the exact set of
reference events used by researchers will vary. It is also possible that
some chronic HCCs are not coded in sequential monitoring periods for
legitimate reasons (e.g.~if a beneficiary does not receive care for the
condition in both periods).

In addition, we note that prior work has found that underreporting is
consistently present in TM, but the degree varies by HCC and by
year\textsuperscript{1,2}. This also means that underreporting estimates
will likely vary depending on researchers' choice of reference events.
However, current coding intensity estimation approaches overwhelmingly
disregard underreporting overall, so our proposed estimator is a helpful
first step towards better accounting for this issue and avoiding
overestimation of coding intensity between Medicare programs. Future
work in undercoding estimation could better account for heterogeneity in
coding behaviors.

\section{Variance and covariance of
estimators}\label{variance-and-covariance-of-estimators}

Variance and (if applicable) covariance are reported by estimator below.

\subsection{\texorpdfstring{\(\hat{F}_s(t)\)}{\textbackslash hat\{F\}\_s(t)}}\label{hatf_st}

\(\hat{F}_s(t)\) is the estimator for event-specific cumulative
incidence. The variance for \(\hat{F}_s(t)\) is
\(\widehat{\text{var}}(\hat{F}_s(t)) = \sum_{t_l \le t} \widehat{\text{var}}(\hat{\theta}(t_l)) + 2 \sum_{t_l < t} \sum_{t_l < t_i \le t} \widehat{\text{cov}}(\hat{\theta}(t_l), \hat{\theta}(t_i))\),
and covariance is
\(\widehat{\text{cov}}(\hat{F}_s(t), \hat{F}_s(u)) = \widehat{\text{var}}(\hat{F}_s(t)) + \sum_{t_l \le t} \sum_{t < t_i \le u} \widehat{\text{cov}}(\hat{\theta}(t_l), \hat{\theta}(t_i))\).
Here \(t_l\) indicates an event reporting time prior to \(t_i\) and
\(u\) is an arbitrary time distinct from \(t\). \(\hat{\theta}(t_i)\)
has variance
\(\widehat{\text{var}}(\widehat{\theta}(t_i))= (\widehat{\theta}(t_i))^2 \times ((r_i - d_{si})/(d_{si}r_i) + \sum_{t_l < t_i} (d_l/(r_l(r_l - d_l))))\)
and covariance
\(\widehat{\text{cov}}(\hat{\theta} (t_i), \hat{\theta} (t_l)) = \hat{\theta} (t_i) \hat{\theta} (t_l)(-(1/r_i) + \sum_{t_l < t_i} (d_l/(r_l(r_l - d_l))))\)\textsuperscript{3},
where \(d_l = \sum d_{s,l}\) is the reported count of events of any
subtype at \(t_l\), \(d_{s,l}\) is analogous to \(d_{s,i}\), and \(r_l\)
is the count of beneficiaries at risk at \(t_l\).

\subsection{\texorpdfstring{\(\hat{\mu}_s(\tau)\)}{\textbackslash hat\{\textbackslash mu\}\_s(\textbackslash tau)}}\label{hatmu_stau}

\(\hat{\mu}_s(\tau)\) is the estimator for
\(\psi = \mu_{s,g=1}(\tau) - \mu_{s,g=0}(\tau)\), or the restricted mean
time lost for event \(s\) up in a single monitoring period with end time
\(\tau\). This estimator has variance
\(\widehat{\text{var}}(\hat{\mu}_s(\tau)) = \sum_{t_i < \tau} (t_{i + 1} - t_i)^2 \widehat{\text{var}}(\hat{F}_s(t_i)) + 2 \sum_{t_i < \tau} \sum_{t_l < t_i} (t_{i+1} - t_i)(t_{l+1} - t_l) \widehat{\text{cov}}(\hat{F}_s(t_i), \hat{F}_s(t_l))\).

\subsection{\texorpdfstring{\(\widehat{\psi}\) and
\(\widehat{\psi}^*\)}{\textbackslash widehat\{\textbackslash psi\} and \textbackslash widehat\{\textbackslash psi\}\^{}*}}\label{widehatpsi-and-widehatpsi}

\(\hat{\psi}\) is the estimand for
\(\psi = \mu_{s,g=1}(\tau) - \mu_{s,g=0}(\tau)\), or the difference in
mean time without event across groups within one monitoring period.
\(\hat{\psi}^*\) is the estimate of the same estimand with
underreporting correction. The variance of \(\widehat{\psi}\) is
\(\widehat{\text{var}}(\widehat{\psi}) = \widehat{\text{var}}(\hat{\mu}_{s,g=1}(\tau)) + \widehat{\text{var}}(\hat{\mu}_{s,g=0}(\tau))\),
as groups are assumed to be independent. It does not change for
\(\widehat{\psi}^*\).

\subsection{\texorpdfstring{\(\widehat{\psi}_M\) and
\(\widehat{\psi}_M^*\)}{\textbackslash widehat\{\textbackslash psi\}\_M and \textbackslash widehat\{\textbackslash psi\}\_M\^{}*}}\label{widehatpsi_m-and-widehatpsi_m}

\(\widehat{\psi}_M\) estimates \(\psi\) across sequential montioring
periods \(\psi_{M} = \psi_m - \psi_{m-1}\), and \(\widehat{\psi}_M^*\)
does the same with an underreporting correction in the comparison group.
The variance of \(\widehat{\psi}_M\) is
\(\widehat{\text{var}}(\widehat{\psi}_M) = \widehat{\text{var}}(\widehat{\psi}_m) + \widehat{\text{var}}(\widehat{\psi}_{m-1})\).
Although the components of this estimator come from the same fixed
population, the correlation between component estimates is assumed to be
negligible because we are comparing differences in restricted mean time
lost across independent groups estimated in disjoint time intervals.
Therefore, covariance is zero. The variance remains unchanged for
\(\widehat{\psi}_M^*\).

\subsection{\texorpdfstring{\(\widehat{\omega}\)}{\textbackslash widehat\{\textbackslash omega\}}}\label{widehatomega}

\(\widehat{\omega}\) estimates possible severity-based upcoding in one
monitoring period. The variance of \(\widehat{\omega}\) is
\(\widehat{\text{var}}(\widehat{\omega}) = \widehat{\text{var}}(\widehat{\omega}_{g=1}) + \widehat{\text{var}}(\widehat{\omega}_{g=0})\),
where
\(\widehat{\text{var}}(\widehat{\omega_g}) = \widehat{\text{var}}(\hat{\mu}_{s=k}(\tau)) + \widehat{\text{var}}(\hat{\mu}_{s=1}(\tau)) - 2\widehat{\text{cov}}(\hat{\mu}_{s=k}(\tau), \hat{\mu}_{s=1}(\tau))\)
and
\(\widehat{\text{cov}}(\hat{\mu}_{s=k}(\tau), \hat{\mu}_{s=1}(\tau)) = \sum_{i} \sum_j \widehat{\text{cov}}(\hat{F}_k(t_{i-1}), \hat{F}_1(t_{j-1}))\)
for distinct event times \(t_i\) and \(t_j\). Here, \(t_i\) corresponds
to event reporting increments in the cumulative incidence function for
severity level \(k\), while \(t_j\) corresponds to the equivalent for
severity level \(1\). Although event reporting times are the same, we
write these using separate variables because covariance is estimated
between all pairs of increments for these two severity levels.

\newpage{}

\section{Summary statistics from All of Us survey
data}\label{sec-allofus}

\begin{longtable}[]{@{}ll@{}}

\caption{\label{tbl-aousurveys}\textbf{Included surveys from All of Us}.
These data represent included survey respondents with at least one
self-reported health condition that mapped to a Hierarchical Condition
Category (HCC) in Version 28 of the Medicare Advantage risk adjustment
algorithm. Only sets of co-occurring HCCs with more than 21 respondents
(which also excludes several sets with more than 20 respondents) were
used in any summary tables to comply with the All of Us Data and
Statistics Dissemination Policy.}

\tabularnewline

\toprule\noalign{}
All of Us Survey Title & Number of Unique Respondents \\
\midrule\noalign{}
\endhead
\bottomrule\noalign{}
\endlastfoot
Overall Health & 156 \\
Personal and Family Health History & 15692 \\

\end{longtable}

\newpage{}

\begin{longtable}[]{@{}ll@{}}

\caption{\label{tbl-num-repondents}\textbf{Number of respondents by HCC
with available survey questions}. These data represent included survey
respondents with at least one self-reported health condition that mapped
to a Hierarchical Condition Category (HCC) in Version 28 of the Medicare
Advantage risk adjustment algorithm. Only sets of co-occurring HCCs with
more than 21 respondents (which also excludes several sets with more
than 20 respondents) were used in any summary tables to comply with the
All of Us Data and Statistics Dissemination Policy.}

\tabularnewline

\toprule\noalign{}
Version 28 HCC & Number of Respondents \\
\midrule\noalign{}
\endhead
\bottomrule\noalign{}
\endlastfoot
HCC1 & 546 \\
HCC20 & 339 \\
HCC21 & 5775 \\
HCC22 & 260 \\
HCC23 & 6772 \\
HCC35 & 156 \\
HCC38 & 2114 \\
HCC51 & 4283 \\
HCC62 & 156 \\
HCC64 & 100 \\
HCC65 & 546 \\
HCC77 & 156 \\
HCC78 & 511 \\
HCC93 & 1131 \\
HCC109 & 1018 \\
HCC155 & 1467 \\
HCC182 & 47 \\
HCC221 & 156 \\
HCC226 & 73 \\
HCC228 & 399 \\
HCC238 & 1635 \\
HCC249 & 373 \\
HCC264 & 62 \\
HCC267 & 123 \\
HCC276 & 156 \\
HCC280 & 273 \\
HCC300 & 587 \\
HCC327 & 31 \\
HCC328 & 110 \\
HCC398 & 808 \\

\end{longtable}

\newpage{}

\begin{longtable}[]{@{}rr@{}}

\caption{\label{tbl-hccspp}\textbf{Distribution of HCC-relevant
conditions per person}. These data represent included survey respondents
with at least one self-reported health condition that mapped to a
Hierarchical Condition Category (HCC) in Version 28 of the Medicare
Advantage risk adjustment algorithm. Only sets of co-occurring HCCs with
more than 21 respondents (which also excludes several sets with more
than 20 respondents) were used in any summary tables to comply with the
All of Us Data and Statistics Dissemination Policy.}

\tabularnewline

\toprule\noalign{}
Number of HCC-Relevant Conditions Per Person & Number of Respondents \\
\midrule\noalign{}
\endhead
\bottomrule\noalign{}
\endlastfoot
1 & 5636 \\
2 & 6657 \\
3 & 3163 \\
4 & 236 \\
5 & 156 \\

\end{longtable}

\newpage{}

\begin{longtable}[]{@{}lrr@{}}

\caption{\label{tbl-age}\textbf{Respondent age at survey response}. The
following tables show respondent information using categories provided
by All of Us. Not all respondents responded to all questions so counts
may vary. The rows in each table correspond to available fields in All
of Us. These characteristics are not displayed jointly because that
would result in sample sizes too small to comply with the All of Us Data
and Statistics Dissemination Policy.}

\tabularnewline

\toprule\noalign{}
Age Group (years) & Number of Respondents & Percent of Respondents
(\%) \\
\midrule\noalign{}
\endhead
\bottomrule\noalign{}
\endlastfoot
{[}65-70) & 5088 & 32 \\
{[}70-75) & 5767 & 36 \\
{[}75-80) & 3383 & 21 \\
{[}80-85) & 1362 & 9 \\
{[}85-90) & 250 & 2 \\

\end{longtable}

\newpage{}

\begin{longtable}[]{@{}
  >{\raggedright\arraybackslash}p{(\linewidth - 4\tabcolsep) * \real{0.5288}}
  >{\raggedleft\arraybackslash}p{(\linewidth - 4\tabcolsep) * \real{0.2115}}
  >{\raggedleft\arraybackslash}p{(\linewidth - 4\tabcolsep) * \real{0.2596}}@{}}

\caption{\label{tbl-sex}\textbf{Respondents' self-reported sex at
birth}. The following tables show respondent information using
categories provided by All of Us. Not all respondents responded to all
questions so counts may vary. The rows in each table correspond to
available fields in All of Us. These characteristics are not displayed
jointly because that would result in sample sizes too small to comply
with the All of Us Data and Statistics Dissemination Policy.}

\tabularnewline

\toprule\noalign{}
\begin{minipage}[b]{\linewidth}\raggedright
Self-Reported Sex at Birth
\end{minipage} & \begin{minipage}[b]{\linewidth}\raggedleft
Number of Respondents
\end{minipage} & \begin{minipage}[b]{\linewidth}\raggedleft
Percent of Respondents (\%)
\end{minipage} \\
\midrule\noalign{}
\endhead
\bottomrule\noalign{}
\endlastfoot
Female & 8461 & 53 \\
Male & 6975 & 44 \\
No matching concept & 269 & 2 \\
Not male, not female, prefer not to answer, or skipped & 143 & 1 \\

\end{longtable}

\newpage{}

\begin{longtable}[]{@{}
  >{\raggedright\arraybackslash}p{(\linewidth - 4\tabcolsep) * \real{0.3288}}
  >{\raggedleft\arraybackslash}p{(\linewidth - 4\tabcolsep) * \real{0.3014}}
  >{\raggedleft\arraybackslash}p{(\linewidth - 4\tabcolsep) * \real{0.3699}}@{}}

\caption{\label{tbl-ethnicity}\textbf{Respondents' self-reported
ethnicity}. The following tables show respondent information using
categories provided by All of Us. Not all respondents responded to all
questions so counts may vary. The rows in each table correspond to
available fields in All of Us. These characteristics are not displayed
jointly because that would result in sample sizes too small to comply
with the All of Us Data and Statistics Dissemination Policy.}

\tabularnewline

\toprule\noalign{}
\begin{minipage}[b]{\linewidth}\raggedright
Self-Reported Ethnicity
\end{minipage} & \begin{minipage}[b]{\linewidth}\raggedleft
Number of Respondents
\end{minipage} & \begin{minipage}[b]{\linewidth}\raggedleft
Percent of Respondents (\%)
\end{minipage} \\
\midrule\noalign{}
\endhead
\bottomrule\noalign{}
\endlastfoot
Hispanic or Latino & 633 & 4 \\
Not Hispanic or Latino & 14442 & 91 \\
Prefer not to answer & 34 & 0 \\
Skipped & 639 & 4 \\
Neither & 100 & 1 \\

\end{longtable}

\newpage{}

\begin{longtable}[]{@{}
  >{\raggedright\arraybackslash}p{(\linewidth - 4\tabcolsep) * \real{0.3467}}
  >{\raggedleft\arraybackslash}p{(\linewidth - 4\tabcolsep) * \real{0.2933}}
  >{\raggedleft\arraybackslash}p{(\linewidth - 4\tabcolsep) * \real{0.3600}}@{}}

\caption{\label{tbl-race}\textbf{Respondents' self-reported race}. The
following tables show respondent information using categories provided
by All of Us. Not all respondents responded to all questions so counts
may vary. The rows in each table correspond to available fields in All
of Us. These characteristics are not displayed jointly because that
would result in sample sizes too small to comply with the All of Us Data
and Statistics Dissemination Policy.}

\tabularnewline

\toprule\noalign{}
\begin{minipage}[b]{\linewidth}\raggedright
Self-Reported Race
\end{minipage} & \begin{minipage}[b]{\linewidth}\raggedleft
Number of Respondents
\end{minipage} & \begin{minipage}[b]{\linewidth}\raggedleft
Percent of Respondents (\%)
\end{minipage} \\
\midrule\noalign{}
\endhead
\bottomrule\noalign{}
\endlastfoot
Asian & 260 & 2 \\
Black or African American & 752 & 5 \\
White & 13347 & 84 \\
Another single population & 49 & 0 \\
More than one population & 129 & 1 \\
I prefer not to answer & 34 & 0 \\
Skipped & 639 & 4 \\
None indicated & 538 & 3 \\
None of these & 100 & 1 \\

\end{longtable}

\newpage{}

\section{\texorpdfstring{Additional information about \texttt{upcoding}
R
package}{Additional information about upcoding R package}}\label{additional-information-about-upcoding-r-package}

This package is available under a MIT license, a copy of which is
included in the package repository. A brief tutorial on specific package
functions is available in the package's Github repository.

\subsection{Baseline data
simulation}\label{sec-baseline-data-simulation}

We simulated co-occurring hierarchical condition categories (HCCs) using
data from the All of Us study\textsuperscript{4}. This study was
designed to enroll a million participants across the United States,
focusing especially on groups historically underrepresented in clinical
and biomedical research\textsuperscript{4} and included questions that
overlapped with many HCCs in the current version of the Centers for
Medicare and Medicaid Services' (CMS) Medicare Advantage (MA) risk
adjustment algorithm (i.e.~CMS-HCC Version 28, or V28). V28 HCCs (listed
in Appendix Table~\ref{tbl-v28hccs}) were manually mapped to Systemized
Nomenclature of Medicine-Clinical Terms (SNOMED-CT) concept identifiers
for survey questions (using version 7 of All of Us data; see this
project's Github repository for mapping). Then, survey responses to all
available SNOMED-CT concepts were queried from the All of Us database.
Included surveys, available respondent sociodemographic characteristics,
and coverage of V28 HCCs are described in Section~\ref{sec-add-info}.

For each surveyed person, self-reported co-occurring V28 HCCs were
extracted. This was then summarized into a table of unique co-occurring
HCC sets and a count of the number of All of Us survey respondents that
reported having these diagnoses. In line with the All of Us Data and
Statistics Dissemination Policy, only sets of co-occurring HCCs with
more than 21 respondents were exported from All of Us's platform (the
Researcher Workbench) to be used in analyses. This both omits all sets
of 20 or fewer respondents and also excludes several sets of
co-occurring HCCs with more than 20 respondents. Users also have the
option to use alternate sets of co-occurring V28 HCCs for baseline data
sampling if they prefer.

\subsection{Upcoding and loss to follow up
simulation}\label{upcoding-and-loss-to-follow-up-simulation}

The package enables users to simulate two types of upcoding realistic to
MA: any-available or severity-based, although the latter can only occur
if an HCC has competing events. This upcoding is randomly split across a
user-specified number of time points. Even though there is inherent
right censoring in upcoded data (because only a proportion of all
simulated individuals will be upcoded or coded at all), the package
provides users with the option to include additional right censoring.
This is meant to be representative of loss to follow up or death, both
of which are known issues in following a population aged 65 years and
older over time\textsuperscript{5}. Users specify the proportion of loss
to follow up they want to include, and a randomly selected set of rows
(e.g., individuals) are right censored over each time point. Once
someone is lost to follow up, they cannot be coded for any HCC in
subsequent time periods.

\subsection{Undercoding simulation}\label{sec-undercoding-simulation}

The package also separately enables undercoding. Here, users specify a
proportion of all coded diagnoses (from simulated baseline data) to
randomly remove across the entire data set. This occurs on a
dataset-wide level because undercoding is a systemic issue in
Traditional Medicare\textsuperscript{1,2} and so we do not assume that
it impacts specific V28 HCCs disproportionately.

\newpage{}

\section{Additional information about version 28 Hierarachical Condition
Categories (HCCs)}\label{sec-add-info}

\begin{longtable}[]{@{}
  >{\raggedright\arraybackslash}p{(\linewidth - 4\tabcolsep) * \real{0.3333}}
  >{\raggedright\arraybackslash}p{(\linewidth - 4\tabcolsep) * \real{0.3333}}
  >{\raggedright\arraybackslash}p{(\linewidth - 4\tabcolsep) * \real{0.3333}}@{}}
\caption{\textbf{Hierarchical Condition Categories (HCCs) included in
the Medicare Advantage risk adjustment algorithm version 28 (V28).} Each
of the HCCs in this table are described as written in the Centers for
Medicare and Medicaid Services' (CMS) 2024 Medicare Advantage Advance
Notice\textsuperscript{6}. Competing events, if applicable, are also
noted from the 2024 CMS Risk Adjustment Model Software and ICD-10
Mappings, specifically the Midyear/Final Model Software, which is in the
SAS language\textsuperscript{7}. If a set of competing events is not
N/A, then, if the HCC in that row is coded, none of the competing event
HCCs can also be coded for billing purposes. The hierarchy described in
the ``Competing Events'' column can also be found in the
\texttt{upcoding} R package.}\label{tbl-v28hccs}\tabularnewline
\toprule\noalign{}
\begin{minipage}[b]{\linewidth}\raggedright
HCC
\end{minipage} & \begin{minipage}[b]{\linewidth}\raggedright
HCC Description
\end{minipage} & \begin{minipage}[b]{\linewidth}\raggedright
Competing Events
\end{minipage} \\
\midrule\noalign{}
\endfirsthead
\toprule\noalign{}
\begin{minipage}[b]{\linewidth}\raggedright
HCC
\end{minipage} & \begin{minipage}[b]{\linewidth}\raggedright
HCC Description
\end{minipage} & \begin{minipage}[b]{\linewidth}\raggedright
Competing Events
\end{minipage} \\
\midrule\noalign{}
\endhead
\bottomrule\noalign{}
\endlastfoot
HCC1 & HIV/AIDS & N/A \\
HCC2 & Septicemia, sepsis, systemic inflammatory response syndrome/shock
& N/A \\
HCC6 & Opportunistic infections & N/A \\
HCC17 & Cancer metastatic to lung, liver, brain, and other organs; acute
myeloid leukemia except promyelocytic & HCC18, HCC19, HCC20, HCC21,
HCC22, HCC23 \\
HCC18 & Cancer metastatic to bone, other and unspecified metastatic
cancer; acute leukemia except myeloid & HCC19, HCC20, HCC21, HCC22,
HCC23 \\
HCC19 & Myelodysplastic syndromes, multiple myeloma, and other cancers &
HCC20, HCC21, HCC22, HCC23 \\
HCC20 & Lung and other severe cancers & HCC21, HCC22, HCC23 \\
HCC21 & Lymphoma and other cancers & HCC22, HCC23 \\
HCC22 & Bladder, colorectal, and other cancers & HCC23 \\
HCC23 & Prostate, breast, and other cancers and tumors & N/A \\
HCC35 & Pancreas transplant status & HCC36, HCC37, HCC38 \\
HCC36 & Diabetes with severe acute complications & HCC37, HCC38 \\
HCC37 & Diabetes with chronic complications & HCC38 \\
HCC38 & Diabetes with glycemic, unspecified, or no complications &
N/A \\
HCC48 & Morbid obesity & N/A \\
HCC49 & Specified lysosomal storage disorders & N/A \\
HCC50 & Amyloidosis, porphyria, and other specified metabolic disorders
& N/A \\
HCC51 & Addison's and Cushing's diseases, acromegaly, and other
specified endocrine disorders & N/A \\
HCC62 & Liver transplant status/complications & HCC63, HCC64, HCC65,
HCC68 \\
HCC63 & Chronic liver failure/end-stage liver disorders & HCC64, HCC65,
HCC68, HCC202 \\
HCC64 & Cirrhosis of liver & HCC65, HCC68 \\
HCC65 & Chronic hepatitis & N/A \\
HCC68 & Cholangitis and obstruction of bile duct without gallstones &
N/A \\
HCC77 & Intestine transplant status/complications & HCC78, HCC80,
HCC81 \\
HCC78 & Intestinal obstruction/perforation & N/A \\
HCC79 & Chronic pancreatitis & N/A \\
HCC80 & Crohn's disease (regional enteritis) & HCC81 \\
HCC81 & Ulcerative colitis & N/A \\
HCC92 & Bone/joint/muscle/severe soft tissue infections/necrosis &
N/A \\
HCC93 & Rheumatoid arthritis and other specified inflammatory rheumatic
disorders & HCC94 \\
HCC94 & Systemic lupus erythematosus and other specified systemic
connective tissue disorders & N/A \\
HCC107 & Sickle cell anemia (Hb-SS) and thalassemia beta zero &
HCC108 \\
HCC108 & Sickle cell disorders, except sickle cell anemia (Hb-SS) and
thalassemia beta zero; beta thalassemia major & N/A \\
HCC109 & Acquired hemolytic, aplastic, and sideroblastic anemias &
N/A \\
HCC111 & Hemophilia, male & HCC112 \\
HCC112 & Immune thrombocytopenia and specified coagulation defects and
hemorrhagic conditions & N/A \\
HCC114 & Common variable and combined immunodeficiencies & HCC115 \\
HCC115 & Specified immunodeficiencies and white blood cell disorders &
N/A \\
HCC125 & Dementia, severe & HCC126, HCC127 \\
HCC126 & Dementia, moderate & HCC127 \\
HCC127 & Dementia, mild or unspecified & N/A \\
HCC135 & Drug use with psychotic complications & HCC136, HCC137, HCC138,
HCC139 \\
HCC136 & Alcohol use with psychotic complications & HCC137, HCC138,
HCC139 \\
HCC137 & Drug use disorder, moderate/severe, or drug use with
non-psychotic complications & HCC138, HCC139 \\
HCC138 & Drug use disorder, mild, uncomplicated, except cannabis &
HCC139 \\
HCC139 & Alcohol use disorder, moderate/severe, or alcohol use with
specified non-psychotic complications & N/A \\
HCC151 & Schizophrenia & HCC152, HCC153, HCC154, HCC155 \\
HCC152 & Psychosis, except schizophrenia & HCC153, HCC154, HCC155 \\
HCC153 & Personality disorders; anorexia/bulimia nervosa & HCC154,
HCC155 \\
HCC154 & Bipolar disorders without psychosis & HCC155 \\
HCC155 & Major depression, moderate or severe, without psychosis &
N/A \\
HCC180 & Quadriplegia & HCC181, HCC182, HCC253, HCC254 \\
HCC181 & Paraplegia & HCC182, HCC254 \\
HCC182 & Spinal cord disorders/injuries & N/A \\
HCC190 & Amyotrophic lateral sclerosis and other motor neuron disease,
spinal muscular atrophy & N/A \\
HCC191 & Quadriplegic cerebral palsy & HCC180, HCC181, HCC182, HCC192,
HCC253, HCC254 \\
HCC192 & Cerebral palsy, except quadriplegic & HCC180, HCC181, HCC182,
HCC253, HCC254 \\
HCC193 & Chronic inflammatory demyelinating polyneuritis and multifocal
motor neuropathy & N/A \\
HCC195 & Myasthenia gravis with (acute) exacerbation & HCC196 \\
HCC196 & Myasthenia gravis without (acute) exacerbation and other
myoneural disorders & N/A \\
HCC197 & Muscular dystrophy & N/A \\
HCC198 & Multiple sclerosis & N/A \\
HCC199 & Parkinson and other degenerative disease of basal ganglia &
N/A \\
HCC200 & Friedreich and other hereditary ataxias; Huntington disease &
N/A \\
HCC201 & Seizure disorders and convulsions & N/A \\
HCC202 & Coma, brain compression/anoxic damage & N/A \\
HCC211 & Respirator dependence/tracheostomy status/complications &
HCC212, HCC213 \\
HCC212 & Respiratory arrest & HCC213 \\
HCC213 & Cardio-respiratory failure and shock & N/A \\
HCC221 & Heart transplant status/complications & HCC222, HCC223, HCC224,
HCC225, HCC226, HCC227 \\
HCC222 & End stage heart failure & HCC223, HCC224, HCC225, HCC226,
HCC227 \\
HCC223 & Heart failure with heart assist device/artificial heart &
HCC224, HCC225, HCC226, HCC227 \\
HCC224 & Acute on chronic heart failure & HCC225, HCC226, HCC227 \\
HCC225 & Acute heart failure (excludes acute on chronic) & HCC226,
HCC227 \\
HCC226 & Heart failure, except end stage and acute & HCC227 \\
HCC227 & Cardiomyopathy/myocarditis & N/A \\
HCC228 & Acute myocardial infarction & HCC229 \\
HCC229 & Unstable angina and other acute ischemic heart disease & N/A \\
HCC238 & Specified heart arrhythmias & N/A \\
HCC248 & Intracranial hemorrhage & HCC249 \\
HCC249 & Ischemic or unspecified stroke & N/A \\
HCC253 & Hemiplegia/hemiparesis & HCC254 \\
HCC254 & Monoplegia, other paralytic syndromes & N/A \\
HCC263 & Atherosclerosis of arteries of the extremities with ulceration
or gangrene & HCC264, HCC383, HCC409 \\
HCC264 & Vascular disease with complications & N/A \\
HCC267 & Deep vein thrombosis and pulmonary embolism & N/A \\
HCC276 & Lung transplant status/complications & HCC277, HCC278, HCC279,
HCC280 \\
HCC277 & Cystic fibrosis & HCC278, HCC279, HCC280 \\
HCC278 & Idiopathic pulmonary fibrosis and lung involvement in systemic
sclerosis & HCC279, HCC280 \\
HCC279 & Severe persistent asthma & HCC280 \\
HCC280 & Chronic obstructive pulmonary disease, interstitial lung
disorders, and other chronic lung disorders & N/A \\
HCC282 & Aspiration and specified bacterial pneumonias & HCC283 \\
HCC283 & Empyema, lung abscess & N/A \\
HCC298 & Severe diabetic eye disease, retinal vein occlusion, and
vitreous hemorrhage & N/A \\
HCC300 & Exudative macular degeneration & N/A \\
HCC326 & Chronic kidney disease, stage 5 & HCC327, HCC328, HCC329 \\
HCC327 & Chronic kidney disease, severe (stage 4) & HCC328, HCC329 \\
HCC328 & Chronic kidney disease, moderate (stage 3B) & HCC329 \\
HCC329 & Chronic kidney disease, moderate (stage 3, except 3B) & N/A \\
HCC379 & Pressure ulcer of skin with necrosis through to muscle, tendon,
or bone & HCC380, HCC381, HCC382, HCC383 \\
HCC380 & Chronic ulcer of skin, except pressure, through to bone or
muscle & HCC381, HCC382, HCC383 \\
HCC381 & Pressure ulcer of skin with full thickness skin loss & HCC382,
HCC383 \\
HCC382 & Pressure ulcer of skin with partial thickness skin loss &
HCC383 \\
HCC383 & Chronic ulcer of skin, except pressure, not specified as
through to bone or muscle & N/A \\
HCC385 & Severe skin burn & N/A \\
HCC387 & Pemphigus, pemphigoid, and other specified autoimmune skin
disorders & N/A \\
HCC397 & Major head injury with loss of consciousness (\textgreater{} 1
hour) & HCC202, HCC398, HCC399 \\
HCC398 & Major head injury with loss of consciousness (\textless{} 1
hour or unspecified) & HCC202, HCC399 \\
HCC399 & Major head injury without loss of consciousness & N/A \\
HCC401 & Vertebral fractures without spinal cord injury & N/A \\
HCC402 & Hip fracture/dislocation & N/A \\
HCC405 & Traumatic amputations and complications & HCC409 \\
HCC409 & Amputation status, lower limb/amputation complications & N/A \\
HCC454 & Stem cell, including bone marrow, transplant
status/complications & N/A \\
HCC463 & Artificial openings for feeding or elimination & N/A \\
\end{longtable}

\newpage{}

\section{Additional simulation results}\label{sec-addl-sim}

\begin{figure}[H]

\centering{

\pandocbounded{\includegraphics[keepaspectratio]{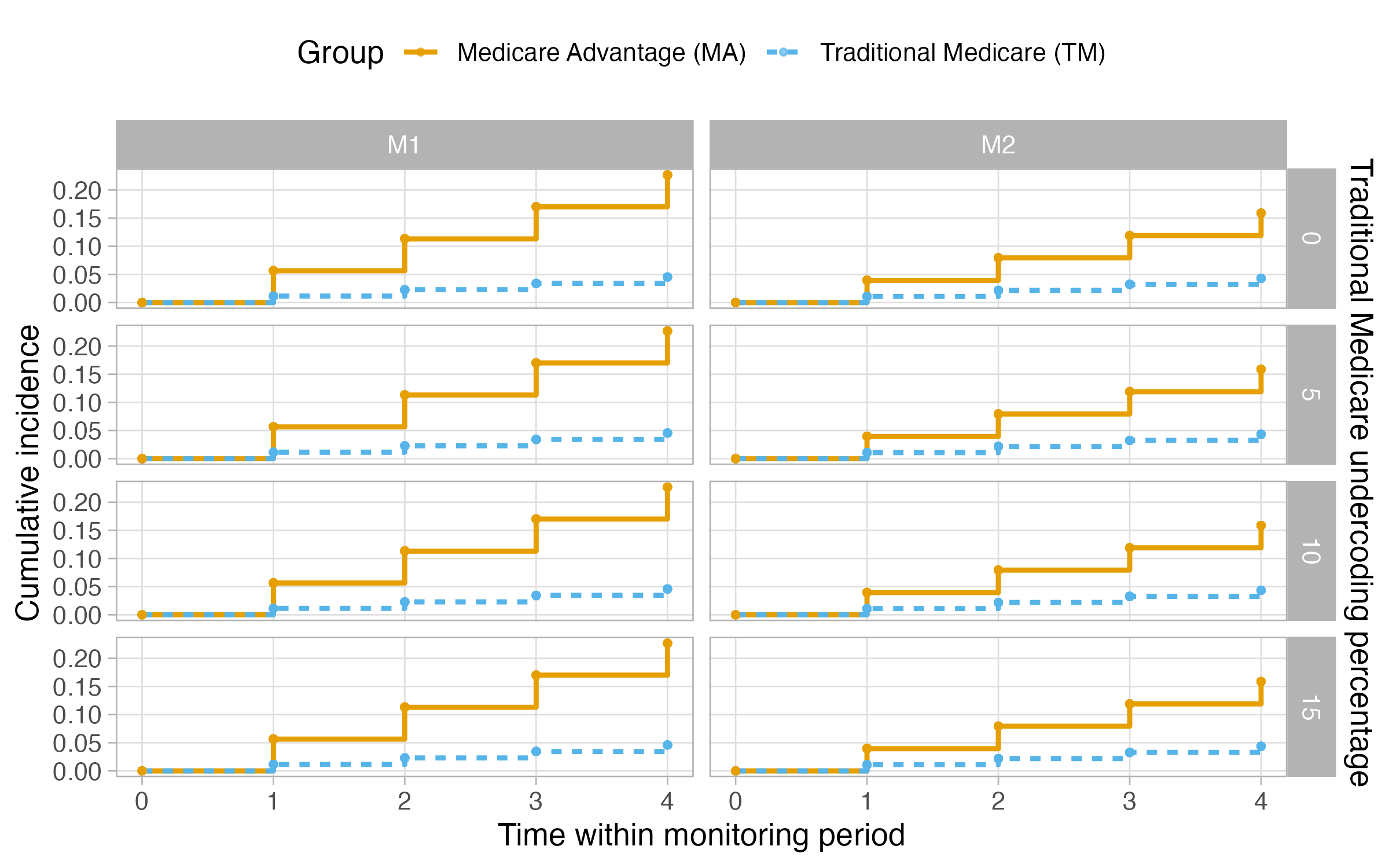}}

}

\caption{\label{fig-supp-2-1}\textbf{Cumulative incidence functions for
the Specified Heart Arrhythmias Hierarchical Condition Category (HCC) in
simulated Medicare Advantage (MA) and Traditional Medicare (TM) groups:
25\% any-available MA upcoding.} Specified heart arrhythmias corresponds
to HCC238, which does not have any competing events. For this HCC, 25\%
of any-available individuals in the MA group are upcoded and 5\% of
any-available individuals in the TM comparison group are upcoded. The
first monitoring period is labeled M1, and the second monitoring period
is labeled M2. Given the large sample size, confidence intervals are
very narrow and are therefore omitted as they cannot be distinguished
visually.}

\end{figure}%

\begin{figure}[H]

\centering{

\pandocbounded{\includegraphics[keepaspectratio]{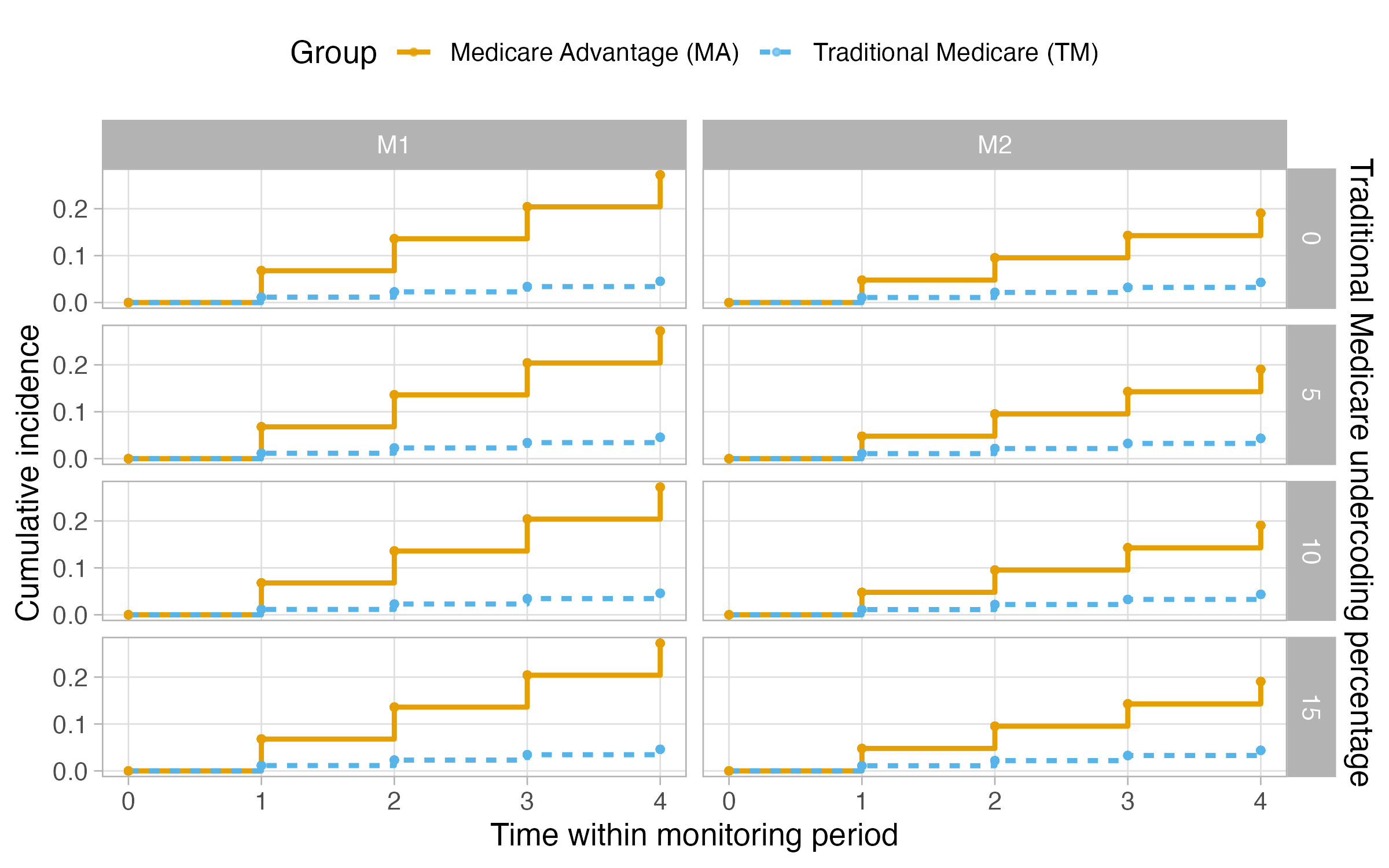}}

}

\caption{\label{fig-supp-2-2}\textbf{Cumulative incidence functions for
the Specified Heart Arrhythmias Hierarchical Condition Category (HCC) in
simulated Medicare Advantage (MA) and Traditional Medicare (TM) groups:
30\% any-available MA upcoding.} Specified heart arrhythmias corresponds
to HCC238, which does not have any competing events. For this HCC, 30\%
of any-available individuals in the MA group are upcoded and 5\% of
any-available individuals in the TM comparison group are upcoded. The
first monitoring period is labeled M1, and the second monitoring period
is labeled M2. Given the large sample size, confidence intervals are
very narrow and are therefore omitted as they cannot be distinguished
visually.}

\end{figure}%

\begin{figure}[H]

\centering{

\pandocbounded{\includegraphics[keepaspectratio]{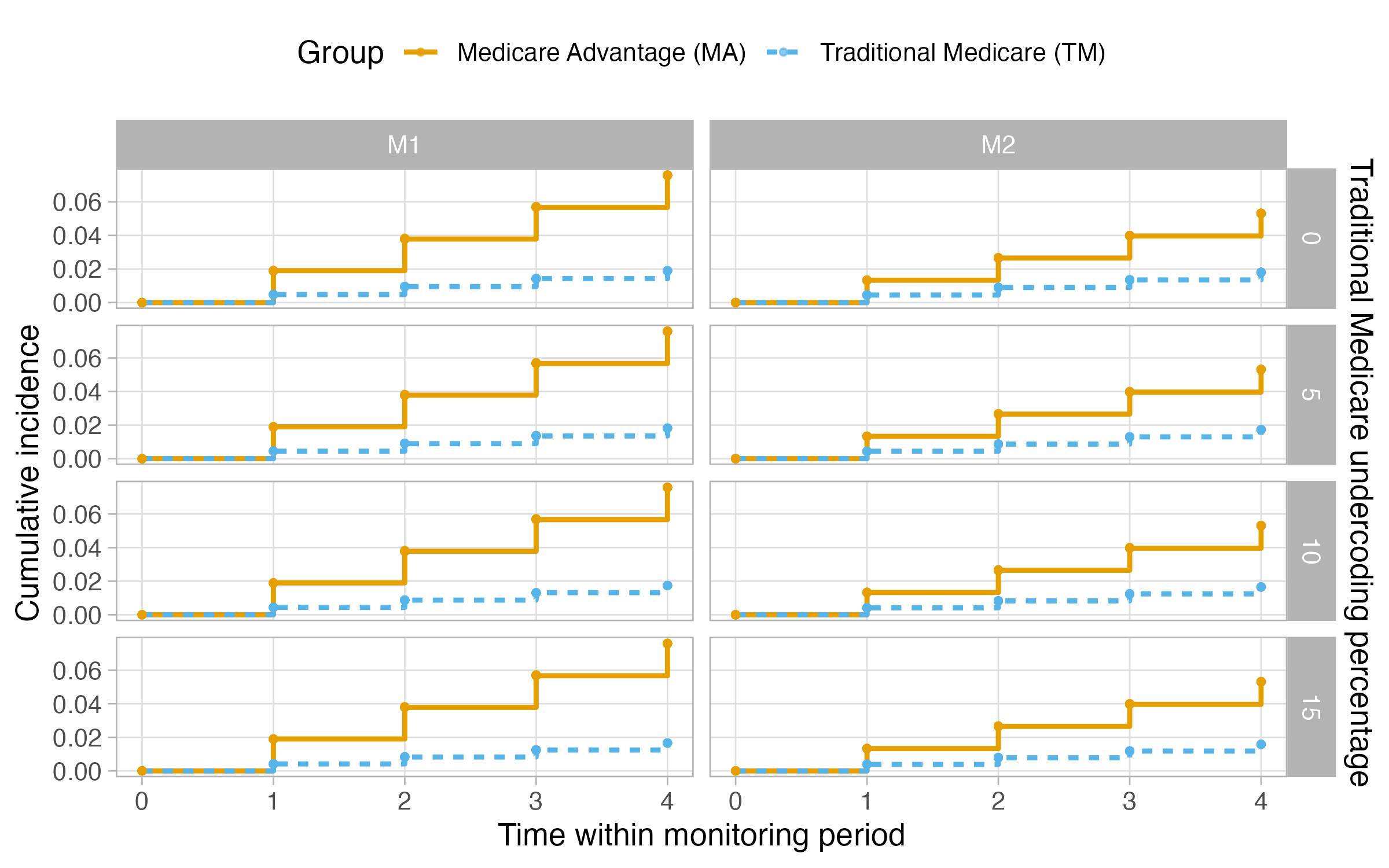}}

}

\caption{\label{fig-supp-2-3}\textbf{Cumulative incidence functions for
the `Dementia, Severe' Hierarchical Condition Category (HCC) in
simulated Medicare Advantage (MA) and Traditional Medicare (TM) groups:
20\% lower severity MA upcoding.} `Dementia, Severe' corresponds to
HCC125, which has lower severity HCCs `Dementia, Moderate' (HCC126) and
`Dementia, Mild or Unspecified' (HCC127). 20\% of any individuals in the
MA group who were previously coded with either HCC126 or HCC127 are
upcoded and 5\% of individuals in the TM comparison group are upcoded
similarly. The first monitoring period is labeled M1, and the second
monitoring period is labeled M2. Given the large sample size, confidence
intervals are very narrow and are therefore omitted as they cannot be
distinguished visually.}

\end{figure}%

\begin{figure}[H]

\centering{

\pandocbounded{\includegraphics[keepaspectratio]{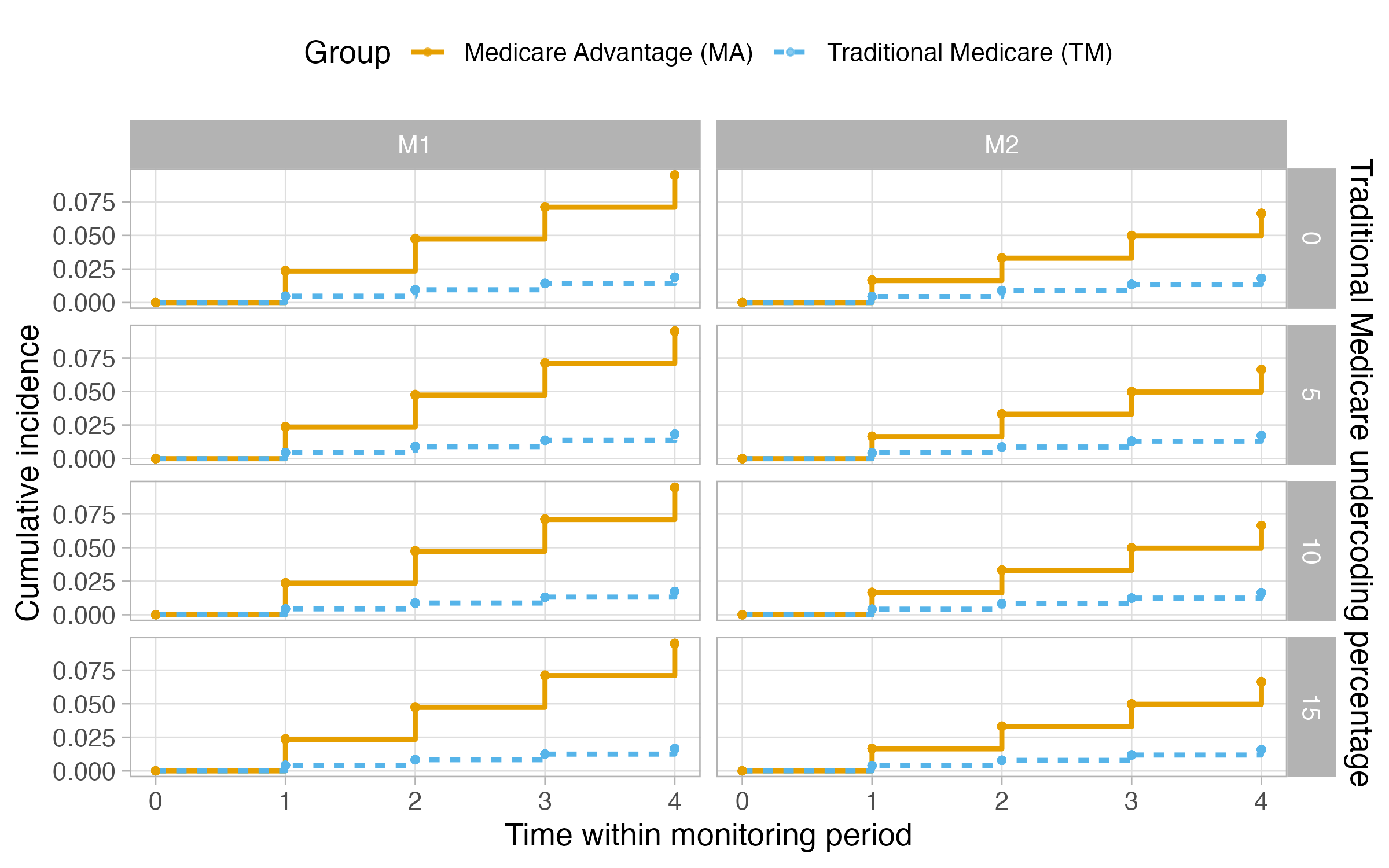}}

}

\caption{\label{fig-supp-2-4}\textbf{Cumulative incidence functions for
the `Dementia, Severe' Hierarchical Condition Category (HCC) in
simulated Medicare Advantage (MA) and Traditional Medicare (TM) groups:
25\% lower severity MA upcoding.} `Dementia, Severe' corresponds to
HCC125, which has lower severity HCCs `Dementia, Moderate' (HCC126) and
`Dementia, Mild or Unspecified' (HCC127). 25\% of any individuals in the
MA group who were previously coded with either HCC126 or HCC127 are
upcoded and 5\% of individuals in the TM comparison group are upcoded
similarly. The first monitoring period is labeled M1, and the second
monitoring period is labeled M2. Given the large sample size, confidence
intervals are very narrow and are therefore omitted as they cannot be
distinguished visually.}

\end{figure}%

\begin{figure}[H]

\centering{

\pandocbounded{\includegraphics[keepaspectratio]{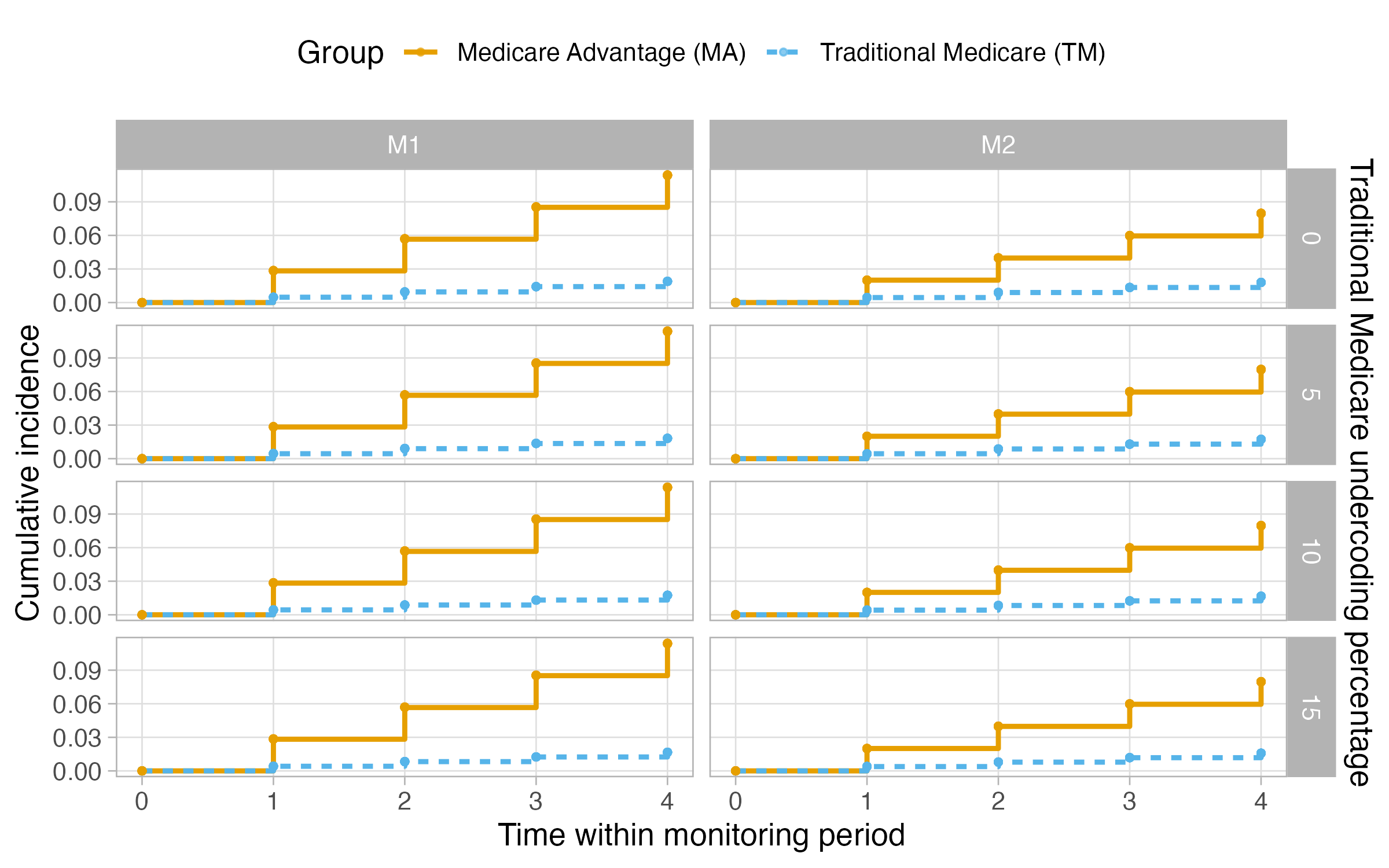}}

}

\caption{\label{fig-supp-2-5}\textbf{Cumulative incidence functions for
the `Dementia, Severe' Hierarchical Condition Category (HCC) in
simulated Medicare Advantage (MA) and Traditional Medicare (TM) groups:
30\% lower severity MA upcoding.} `Dementia, Severe' corresponds to
HCC125, which has lower severity HCCs `Dementia, Moderate' (HCC126) and
`Dementia, Mild or Unspecified' (HCC127). 30\% of any individuals in the
MA group who were previously coded with either HCC126 or HCC127 are
upcoded and 5\% of individuals in the TM comparison group are upcoded
similarly.The first monitoring period is labeled M1, and the second
monitoring period is labeled M2. Given the large sample size, confidence
intervals are very narrow and are therefore omitted as they cannot be
distinguished visually.}

\end{figure}%

\begin{figure}[H]

\centering{

\pandocbounded{\includegraphics[keepaspectratio]{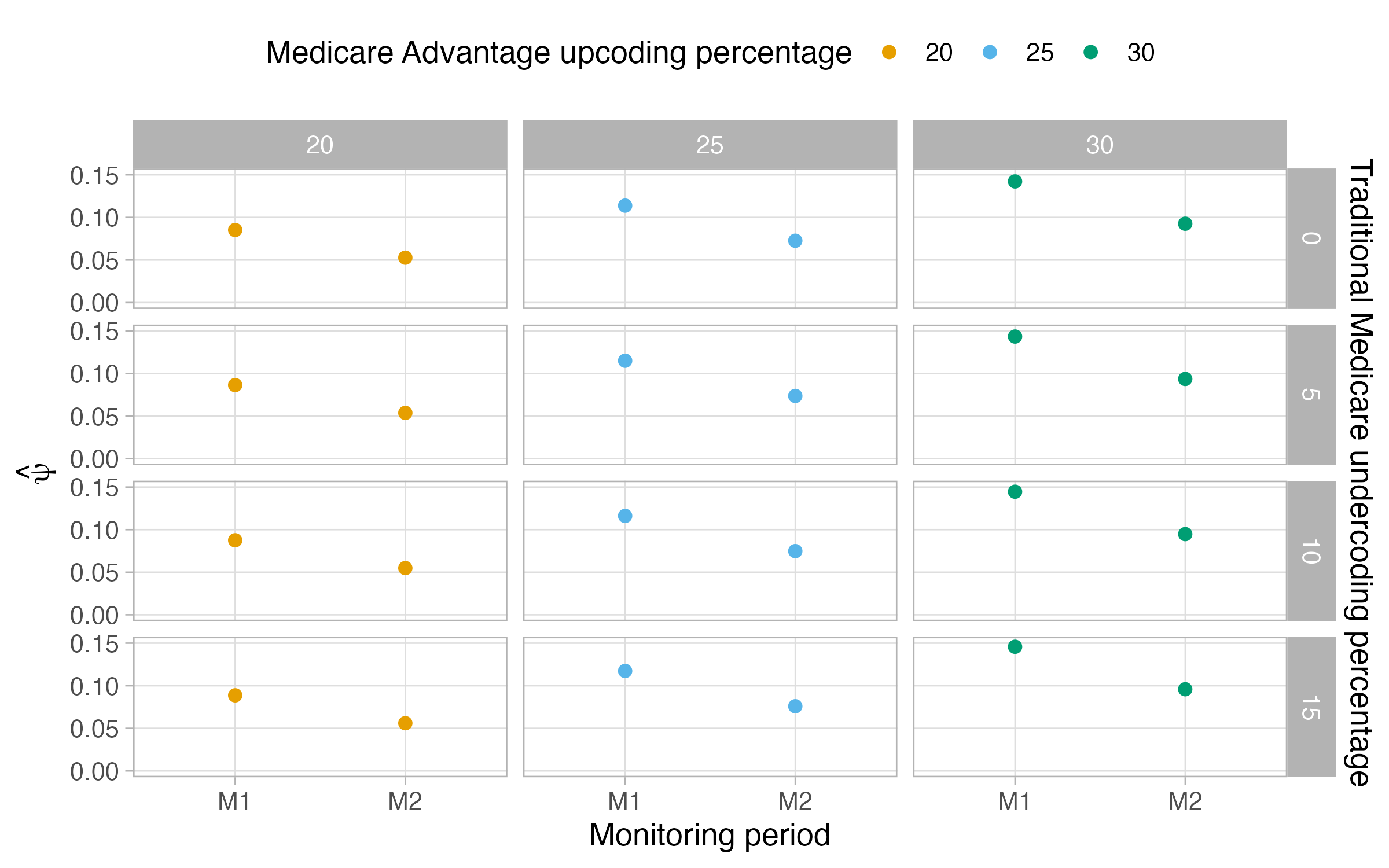}}

}

\caption{\label{fig-supp-2-6}\textbf{Within-monitoring period}
\(\boldsymbol{\psi}\) \textbf{estimates for the `Dementia, Severe'
Hierarchical Condition Category (HCC) in simulated Medicare Advantage
(MA) and Traditional Medicare (TM) groups.} \(\widehat{\psi}\)
corresponds to the proposed estimator for \(\psi\), or the difference in
mean time without event (i.e.~incident coding of the HCC) across groups.
`Dementia, Severe' corresponds to HCC125, which has lower severity HCCs
`Dementia, Moderate' (HCC126) and `Dementia, Mild or Unspecified'
(HCC127). For this HCC, individuals in the MA group who were previously
coded with either HCC126 or HCC127 are upcoded to varying degrees and
individuals in the TM comparison group are upcoded 5\% similarly. The
first monitoring period is labeled M1, and the second monitoring period
is labeled M2. Given the large sample size, confidence intervals are
very narrow and are therefore omitted as they cannot be distinguished
visually.}

\end{figure}%

\newpage{}

\textbf{References}

\phantomsection\label{refs}